\documentclass[12pt,two side ]{report}
\usepackage[utf8]{inputenc}
\usepackage{geometry}
\usepackage{setspace}

\makeatletter
\newif\if@restonecol
\makeatother

\usepackage[linesnumbered,ruled,vlined]{algorithm2e}
\usepackage{algpseudocode}
\usepackage{framed}
\usepackage{amsmath,amssymb,amsfonts}
\usepackage{caption}
\usepackage{graphicx, subfigure}
\usepackage{underscore}
\usepackage{biblatex}
\usepackage{mathrsfs}
\usepackage[nottoc]{tocbibind}
\usepackage{booktabs}

\begin{document}

\begin{titlepage}
    \begin{center}
        \vspace*{1cm}
        
        \Huge
        \textbf{BanditMF: Multi-Armed Bandit Based Matrix Factorization Recommender System}
        
        \vspace{0.5cm}
        \LARGE

        \vspace{0.5cm}
        by
        
        \vspace{1.5cm}
        \textbf{Shenghao XU}\\

        \vfill


        \vspace{0.8cm}

        \Large
        May. 2021\\

    \end{center}
\end{titlepage}
\chapter*{Abstract}
Multi-armed bandits (MAB) provide a principled online learning approach to attain the balance between exploration and exploitation. Generally speaking, in a multi-armed bandit problem, to obtain a higher reward, the agent must choose the optimal action in various states based on previous experience (\textit{exploit}) known actions to obtain a higher score; to discover these actions, the necessary discovery is required (\textit{exploration}). Due to the superior performance and low feedback learning without the learning to act in multiple situations, multi-armed bandits are drawing widespread attention in applications ranging from recommender systems. Likewise, within the recommender system, collaborative filtering (CF) is arguably the earliest and most influential method in the recommender system. The meaning of collaboration is to filter the information through the relationship between the users and the feedback of the user's rating of the items together to find the target users’ preferences. Crucially, new users and an ever-changing pool of recommended items are the challenges that recommender systems need to address. For collaborative filtering, the classical method is to train the model offline, then perform the online testing, but this approach can no longer handle the dynamic changes in user preferences, which is the so-called \textit{cold start}. So, how to effectively recommend items to users in the absence of effective information?

To address the aforementioned problems, a multi-armed bandit based collaborative filtering recommender system has been proposed, named BanditMF. BanditMF is designed to address two challenges in the multi-armed bandits algorithm and collaborative filtering: (1) how to solve the cold start problem for collaborative filtering under the condition of scarcity of valid information, (2) how to solve the sub-optimal problem of bandit algorithms in strong social relations domains caused by independently estimating unknown parameters associated with each user and ignoring correlations between users.

\tableofcontents

\chapter{Introduction}
In this section, first, a detailed account of the recommender system is given, followed by the general information about multi-armed bandits (MAB) and collaborative filtering (CF), which will be introduced. The problem statement will also be covered in this chapter. Also, the aims and contributions of this project will have to be indicated.

\section{General Introduction}
Nowadays, with the rapid development of social networks, e-commerce, and sharing economies, it has become a core link of Internet services to discover users' needs, understand users' behaviors, and screen out the most relevant information and products for users. There is a huge amount of information on the Internet: YouTube users upload over 400 hours of video every minute; WeChat Moments receive 10 billion clicks per day; Instagram Stories have 500 million active users per day\footnote{Data sources: Instagram Revenue and Usage Statistics (2021).}. The recommendation system has become an essential part of various video websites and e-commerce websites to provide various personalized services for users to interact with the system.

Early recommendation systems relied more on simple models or algorithms guided by intuition. For example, the recommendation system based on information retrieval treats the user's information as a \textit{query} and baesd on various types of information to express the item to be recommended as a \textit{document}. Thus, the problem of recommending the most relevant set of items is transformed into the problem of finding the most relevant documents in information retrieval.
\begin{figure}[htbp]
\centering
\includegraphics[scale=0.7]{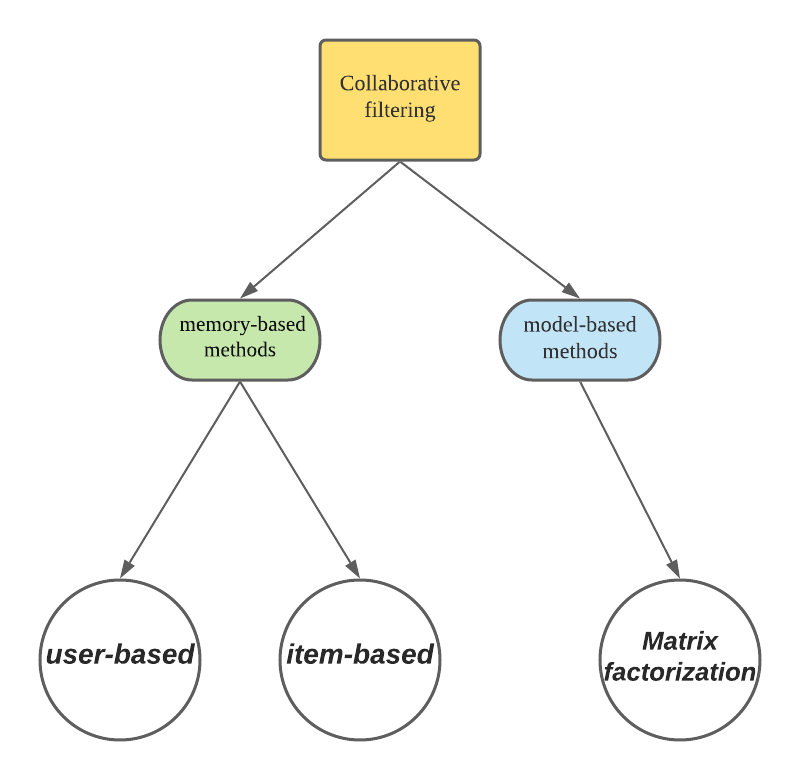}
\caption{Relation diagram of collaborative filtering in recommender systems}
\end{figure}
As the field of recommendation systems has evolved, traditional recommendation systems have been divided into two main categories: content-based filtering (CBF) and collaborative filtering (CF) \cite{cf1999}. The content-based approach describes the nature of users and items by defining explicit attributes (usually defined by experts with professional knowledge) and then recommends items that are \textit{``compatible"} with the user's nature. Traditional collaborative filtering, as shown in Figure 1.1, is categorized into two categories \cite{1423975}: memory-based methods and model-based methods. Item-based collaborative filtering and user-based collaborative filtering fall under the category of memory-based methods. For matrix factorization (MF), it is the most popular collaborative filtering technique that belongs to model-based methods \cite{convMF}.The three methods mentioned above will be elaborated and analyzed in the next chapter.

Matrix factorization attracts a ton of attention due to its good scalability and ease of implementation. In \cite{mf1}, the authors comprehensively introduce the matrix factorization algorithm, which integrates machine learning technology. Other than that, the authors in \cite{convMF} proposed a method that combines matrix factorization with convolutional neural networks (CNN). Matrix factorization is scalable; it can be seamlessly integrated with mainstream models and can also do feature fusion with multiple information sources. So, in \cite{mf2}, a model that fusion the text comment information, geographic neighborhood information, project category information, and popularity information has been proposed to promote the precision of rating prediction. However, these recommender systems focus on how to improve the accuracy of algorithms for a single problem (e.g., rating prediction for matrix factorization), thus neglecting the timeliness and systematicness of human-machine interaction, which makes it difficult to model the vagaries of user behavior and the rapidly changing external environment completely. These matrix factorization-based recommendation systems only exploit the trained model for the recommendation and do no proper exploration. Thus, these recommendation systems tend to produce cookie-cutter results.

In order to achieve a dynamic balance between the system and the user and to shift away from viewing the recommendation system as a running fragment of one recommendation result at a time. Some researchers, such as the authors in \cite{context} have started to explore how to apply some concepts of reinforcement learning to the scenario of recommendation systems. Reinforcement learning often represents a \textit{``feedback system"} in which the user and the system interact to achieve the optimization of the target. The interaction process between the system and users is no longer a \textit{``one-shot deal"}, which is replaced by the dynamic balance between the system and users that achieved through a series of results on the dimension of time.  A number of algorithms have been proposed based on the context-free bandit and the contextual bandit. Further detailed concepts and algorithms of context-free bandit and contextual bandit will be conducted in chapter 2.

However, in the more complex scenario of online mass recommendation systems, the above algorithm has not been favored. Although such algorithms have beautiful proofs and mathematical properties in theory, there are often insurmountable barriers to product experience in practice. As the authors affirm in \cite{conversation}, the contextual bandit algorithm requires a tremendous amount of exploration, resulting in a slow speed. Most of these algorithms are not familiar with the user in the initial period, so the algorithm tends to explore more items to understand the user. Therefore, during this period, the user may be confronted with a number of items that are not as relevant. In these algorithms, the assumption is that once the exploration period is over, the algorithm will be able to learn the user's preferences better and provide more \textit{``trustworthy"} recommendations. However, this hypothetical scenario does not hold in reality. In the real world, only highly loyal users may have the tolerance to accept less relevant recommendations. From the perspective of new users or users with low loyalty, the frequency of use of the product is not high, and when these users find some irrelevant recommendation results in a few uses, it is likely to cause new users to abandon the product permanently.

As alluded to above, collaborative filtering is the process of sifting and filtering information together with the relationship between users and their feedback on the items, to recommend items of interest to the target users. Since coordinated filtering focuses more on judging the similarity between historical records and a user’s similarity to make recommendations, the result is that the content universe of collaborative filtering is almost static and loses the ability to recommend new items. Unfortunately, in many recommendation scenarios, where the universe of content experiences rapid turnover and the popularity of content changes over time. Furthermore, the cold start \cite{cold} problem is also an insurmountable chasm for collaborative filtering.  In former research \cite{difficcf}, it was demonstrated that these problems led to poor performance and difficulty in applying traditional collaborative filtering-based recommender systems.

To address the aforementioned problems, in this work, BanditMF has been proposed, which is a multi-armed bandit based matrix factorization recommender system. This system combines the matrix factorization (MF) which is model-based collaborative filtering with the multi-armed bandit algorithm. BanditMF contains an offline subsystem focusing on matrix factorization and an online subsystem with a multi-armed bandit algorithm as its core. With the power of the multi-armed bandit algorithm, BanditMF solves the cold-start problem of the collaborative filtering method and gives the ability to recommend new items to users. Meanwhile, the bandit algorithm in the online subsystem accepts parameters from the matrix factorization in the offline subsystem to make recommendations. Since the matrix factorization takes into account the connections and similarities between users, the multi-armed bandit algorithm uses these parameters to reduce the irrelevance of the recommended items when exploring, thus reducing the loss of new users with low loyalty due to irrelevant recommendations from the bandit algorithm during the exploration period. The detailed discussion of BanditMF will be introduced in chapter 3.

The remainder of this report is organized as follows: In Chapter 2, the representative multi-armed bandit and collaborative filtering methods will be reviewed and formulated to solve the problem. In Chapter 3, two traditional systems will be illustrated: a collaborative filtering-based recommender system and a hybrid recommender system. Then, based on those traditional systems, BanditMF will be introduced in detail. In chapter 4, the two traditional systems are implemented and experimentally evaluated with BanditMF. We propose the future work and the conclusion of the entire work will be present in chapter 5.
\chapter{Formulation}
In this chapter, context-free bandit and contextual bandit will be introduced, followed by a brief review of the collaborative filtering method.
\section{Multi-Armed Bandit}
The Multi-armed bandits (MAB) framework provides principled solutions for the dilemma between exploration and exploitation.

As one of the classic problems of reinforcement learning—Multi-armed bandits (MAB)—is different from machine learning (e.g., supervised learning), it is based on environmental interaction and trial-and-error to achieve the effect of learning. Such a learning mode is more similar to the human learning mode.

\begin{table}[htbp]
\centering
 
\resizebox{\textwidth}{8mm}{\begin{tabular}{lll}
                                                        & \textbf{Action don't change state of world}      & \textbf{Action change state of the world  }           \\ \cline{2-3} 
\multicolumn{1}{l|}{\textbf{Learn model of outcomes}}            & \multicolumn{1}{l|}{Multi-armed bandits} & \multicolumn{1}{l|}{Reinforcement Learning}  \\ \cline{2-3} 
\multicolumn{1}{l|}{\textbf{Given model of stochastic outcomes}} & \multicolumn{1}{l|}{Decision theory}    & \multicolumn{1}{l|}{Markov Decision Process} \\ \cline{2-3} 
\end{tabular}}
\caption{Four scenarios when making decisions under uncertainty}     

\end{table}

In Table 2.1\footnote[2]{Table from the paper: A Survey on Contextual Multi-armed Bandits.}, four different scenarios have been described, which are the scenarios when making decisions under uncertainty. In the setting of multi-armed bandits, the outcomes (rewards) are unknown, and the outcomes can be stochastic or adversarial. The state of the world can not be changed by actions.

The reinforcement learning process consists of five major elements:  \textit{agent}, \textit{environment}, \textit{state}, \textit{action}, and \textit{reward}. At a certain instant, the intelligence is in a certain state, and after executing an action, the environment receives the action, prompting the agent to move to the next state while feeding back the reward. The agent's purpose is to maximize the accumulated reward and adjust the action accordingly to the amount of the reward to get a higher reward.

For reinforcement learning can using Markov decision process (MDP) to formalize \cite{98sutton}. Formally, a Markov decision process is a 4 tuple (\textit{S}, \textit{A}, \textit{$P_{a}$}, \textit{$R_{a}$}), where:
\begin{itemize}
    \item \textit{S} is denote the set of states called the state space, where \textit{S}={\textit{$S_{1}$}, \textit{$S_{2}$}, \textit{$S_{3}$}, \textit{$S_{4}$},$\cdot \cdot \cdot$}
    
    \item \textit{A} is denote the set of actions called the action space, where \textit{A}={\textit{$A_{1}$}, \textit{$A_{2}$}, \textit{$A_{3}$}, \textit{$A_{4}$},$\cdot \cdot \cdot$}
    
    \item \textit{$P_{a}$} is denote the probability, $P_{a}\left(s, s^{\prime}\right)=\operatorname{Pr}\left(s_{t+1}=s^{\prime} \mid s_{t}=s, a_{t}=a\right)$ is the probability from state $s$ to sate $s^{\prime}$.
    
    \item \textit{$R_{a}$} represent the reward. where $R_{a}\left(s, s^{\prime}\right) =\operatorname{E}\left(r_{t} \mid s_{t}=s, a_{t}=a\right)$.
\end{itemize}

Multi-armed bandits is a classic reinforcement learning problem (essentially a simplified class of reinforcement learning problems) that has non-associative states (learning from only one situation at a time, losing or winning) and focuses only on evaluable feedback. Suppose there is a bandit with \textit{K} arms. Each arm will have a certain probability of getting a reward, so that we have \textit{K} possible actions (each arm corresponds to an action), and the result of each action is only associated with the current state and is not affected by the results of historical actions. This means the reward of each arm is only related to the probability set by the bandit. The previous winning and losing results will not affect this action. So we can formally define multi-armed bandits problems as Markov decision processes with a single state as follows:

\begin{itemize}
    \item No sate \textit{S}, but \textit{$S_{0}$} can denote as hypothetical constant state.
    
    \item \textit{K} arms correspond to \textit{K} different actions, where \textit{A}={\textit{$a_{1}$}, \textit{$a_{2}$}, \textit{$a_{3}$}, ,$\cdot \cdot \cdot$,\textit{$a_{K}$}}.
    
    \item Do not have probability \textit{P}, but it can be assumed that the state always transitions from \textit{$S_{0}$} to \textit{$S_{0}$}.
    
    \item Have a reward function \textit{R} dependence on action \textit{A} instead of state \textit{S}.
\end{itemize}

The key idea of the multi-armed bandits is to make the optimal decisions to maximize the total reward through exploration and exploitation.

\subsection{Context-Free Bandit}
Bandit algorithms can be categorized into \textit{context-free} bandit and \textit{contextual} bandit depending on whether or not contextual features are taken into account. For the traditional \textit{K-armed} bandit due to:
\begin{itemize}
    \item [1.] The arm set $\mathcal{A}_{t}=\left \{ a_{1},a_{2},\dots,a_{k}  \right \}$ with $K$ actions are fixed for all $t\in T$, where $T$ denote the total rounds.
    \item [2.] Both user $u_{t}$ and context $\mathbf{x}_{t, a}$ are unchanged for all $t\in T$.
\end{itemize}
Whenever the arm is pulled, as arm set $\mathcal{A}_{t}$ and context  $\mathbf{x}_{t, a}$ are constant, therefore they do not affect the algorithm. So, for this kind of bandit algorithm, we refer to them as the \textit{context-free} bandit \cite{context}.

For the above context-free bandit, we formulated as follows:
\begin{framed} 
\textbf{Problem Formulation}: The  \textit{context-free} bandit\\
\rule{\textwidth}{0.1mm}
\textbf{Given}: $T$ rounds, $K$ actions, 1 agent.\\
In each round, for $t\in [T]$, agent chooses the policy :
\begin{itemize}
    \item [1.] Policy choose action $a_{t}$.
    \item[2.] For the chosen action, policy observes the rewards $r_{t}$, with $r_{t}\in [0,1]$.
\end{itemize}
\end{framed} 

For context-free bandit, $\mu $ denoting the mean reward vector, where $\mu \in \left [ 0,1 \right ]^{K}$. The mean reward of action $a$ is expressed as $\mu(a) = \mathbb{E}(\mathcal D_{a})$. The $\mu^{*}$ is used to define the best mean reward, where $\mu^{*}:= max_{a_{i}\in \left \{ a_{i} \right \}_{i=1}^{K}}\mu (a)$. We use the notation $\Delta (a)$ to represent the $gap$ of action a. The \textit{gap} describes the difference between the reward received for action $a$ and the reward received for the optimal action, that $\Delta (a):= \mu ^{*}-\mu (a)$ and $a^{*}$ is used to denote the optimal action, which action with $\mu(a)=\mu^{*}$. Formally, \textit{regret} define as follows:
\begin{equation}
    R(T)\triangleq \sum_{t=1}^{T}(\mu^{*}-\mathbb{E}[\mu_{(a_{t})}])= T\cdot \mu^{*}-\mathbb{E}\left [ \sum_{t=1}^{T}\mu_{(a_{t})} \right ]
\end{equation}

The $R(T)$ is called as $regret$ at round $T$, where $a_{t}$ is denote the action chosen at round $t$.
\subsection{Contextual Bandit}
As opposed to the \textit{context-free} bandit, the \textit{contextual} bandit is based on contextual information of the user and items (i.e. feature vector $\mathbf{x}_{t, a}$) to inferring the conditionally expected reward of an action.

For the above contextual bandit, we formulated as follows:
\begin{framed} 
\textbf{Problem Formulation}: The  \textit{contextual} bandit\\
\rule{\textwidth}{0.1mm}
\textbf{Given}: $T$ rounds, users $u_{t}$, arm set $\mathcal{A}_{t}$, feature vector (context) $\mathbf{x}_{t, a}$.\\
In each round, for $t\in [T]$, do :
\begin{itemize}
    \item [1.] For all $a \in \mathcal{A}_{t}$, algorithm observers the current user $u_{t}$ and arm set $\mathcal{A}_{t}$ with correspond feature vector $\mathbf{x}_{t, a}$.
    \item[2.] Algorithm chosen action $a_{t}$, and revives the rewards $r_{t,a_{t}}$.
    \item[3.] With new observation $\mathbf{x}_{t, a}$, $a_{t}$, and $r_{t,a_{t}}$, the algorithm improves its arm selection strategy $\pi$.
\end{itemize}
\end{framed} 

As the variant of the multi-armed bandits, the contextual multi-armed bandits (CMAB) are different from the stochastic bandit and adversarial bandits. In the setting of the stochastic setting, the reward of an arm is sampled from the unknown distribution, or so-called reward distribution. In an adversarial environment, the reward for one arm is chosen by the adversarial, rather than drawn from any distribution. Conversely, under the setting of the contextual bandits, what we are interested in is the situation that we need to observe side information at each time slot, where side in formation is the so-called contexts $\mathbf{x}_{t, a}$. For the contextual bandit, there is a set of actions $\mathcal{A}_{t}$, and each action maps a context to an arm, where at each iteration, before the arm selection, the agent will observe the \textit{d}-dimensional context, where $\mathbf{x}_{t, a}\in \mathbb{R}^{d}$. Based on this context, along with the rewards of arms played in the past, to select the arms to be chosen in the current iteration. The goal of the algorithm is to maximize the reward over finite times $t\in [T]$. $\sum_{t=1}^{T} r_{t, a_{t}}$ denote the total rewards of the algorithm and define the optimal expected cumulative reward as $\mathbb{E}\left [ \sum_{t=1}^{T} r_{t, a_{t}^*} \right ]$, where $a_{t}^*$ denote the optimal arm at time $t$. The \textit{regret} of the contextual bandit algorithm is defined as:
\begin{equation}
    R(T)\triangleq\mathbb{E}\left [ \sum_{t=1}^{T} r_{t, a_{t}^*} \right ]-\mathbb{E}\left [ \sum_{t=1}^{T} r_{t, a_{t}} \right ]
\end{equation}
To address the general situation, different algorithms are proposed, which include LINUCB proposed at \cite{context}, Contextual Thompson Sampling (CTS) described in \cite{tps}, and Neural Bandit in \cite{allesiardo2014neural}, in these algorithms, a linear dependency between the expected payoff of an action and its context is generally hypothesized.
\section{Collaborative Filtering}
In this section, two categories of methods involved in collaborative filtering that will be discussed: memory-based methods and model-based methods.
\subsection{Memory-Based Methods}
Collaborative filtering takes into account the fact that there is a relationship between the items and the user's preference. In memory-based approaches, as the name suggests, a memory-based recommender system typically needs to leverage the entire user-item dataset and make recommendations accordingly.
\begin{figure}[htbp]
\centering
\includegraphics[scale=0.7]{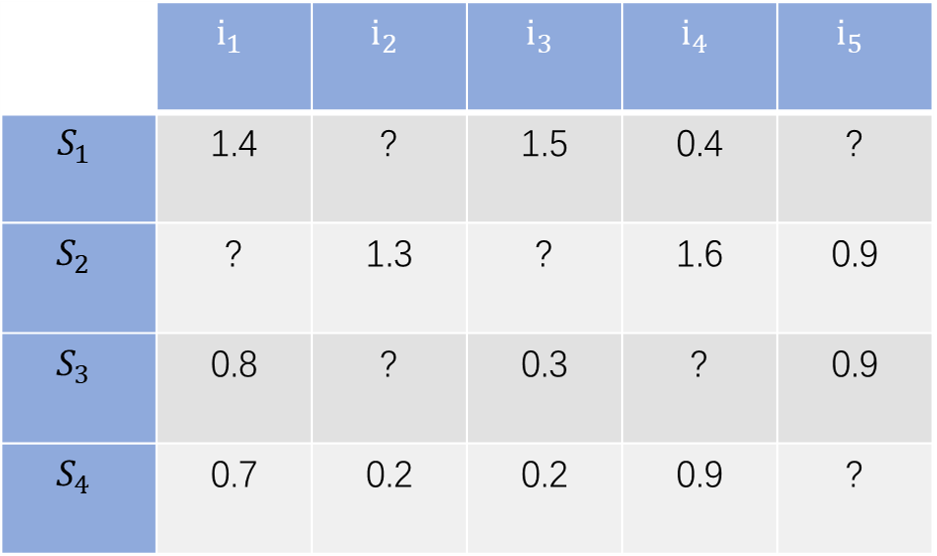}
\caption{User$-$item rating matrix}
\end{figure}
Once we have the user rating data for the items, we can convert them into the form of the matrix as shown in Figure 2.1. The horizontal coordinates are the users, and the vertical coordinates are the items. The corresponding number is the user's rating of the item. The question mark within the matrix indicates that the user has not rated the item, and our goal is to use collaborative filtering to predict unknown ratings. Memory-based collaborative filtering to make recommendations by similar users or similar items, then the question arises how to define similar users or similar items? The answer is similarity. After we have the user-item rating matrix, as shown above, the row vectors in the matrix represent the preferences of each user, and the column vectors represent the attributes of each item. This can be interpreted as meaning that there exists an attribute of the item that attracts users to provide a corresponding rating. So, by computing the similarity between row vectors, we will get the user similarity, and by computing the similarity of vertical vectors, the item similarity will get. \textit{Cosine similarity}, \textit{Pearson correlation coefficient}, and \textit{Euclidean distance} are the common methods to compute the similarity.

This chapter has demonstrated the concept of memory-based collaborative filtering. It is now necessary to explain the two basic approaches of memory-based collaborative filtering: the user-based method and the item-based method.
\subsubsection{User-Based Collaborative Filtering}
For user-based collaborative filtering, in each recommendation, we have a target user who will be recommended. The user-based collaborative filtering algorithm first identifies users that are similar to that particular target user, i.e., users that share the target user's rating pattern. User-based collaborative filtering predicated on similarities including history, preferences, and choices made by users when purchasing, viewing, or certain other behaviors. Based on the ability conferred by similarity, we can estimate the ratings that a target user might give based on similar users' ratings of items that the target user did not give a rating on. Based on the predicted rating, the user-based collaborative filtering algorithm can recommend items with higher ratings to the target users.
\begin{figure}
\centering
\includegraphics[scale=0.5]{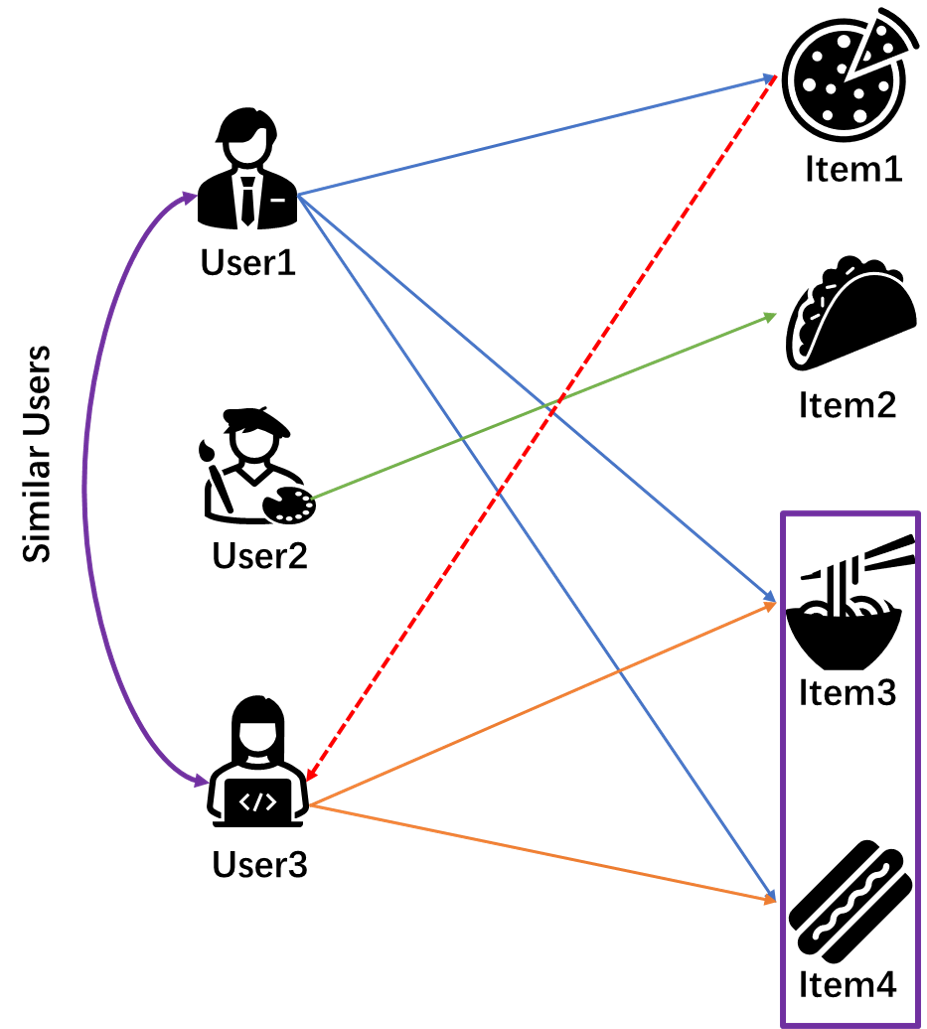}
\caption{User-based collaborative filtering}
\end{figure}

User-based collaborative filtering can be illustrated briefly in Figure 2.2. Recall that the recommendation of the user-based method is based on similar users who share the same preference with the target user. As the example shown in Figure 2.2. Both \textit{user1} and \textit{user3} show a preference for \textit{item3} and \textit{item4}, and for this reason, we consider \textit{user1} and \textit{user3} to be similar users. Then, the user-based collaborative filtering algorithm will recommend \textit{item1} to \textit{user3} that \textit{item1} is the item that wins the positive rating by \textit{user3}’s similar user-\textit{user1}.

\subsubsection{Item-Based Collaborative Filtering}
Item-based collaborative filtering and user-based collaborative filtering are essentially the same. The only difference is that item-based collaborative filtering is based on the similarity between items instead of the similarity between users, as in user-based collaborative filtering. The idea of item-based collaborative filtering is to find items similar to those you have purchased or browsed, then recommend those items to you, i.e., ``Birds of a feather flock together."
\begin{figure}[htbp]
\centering
\includegraphics[scale=0.5]{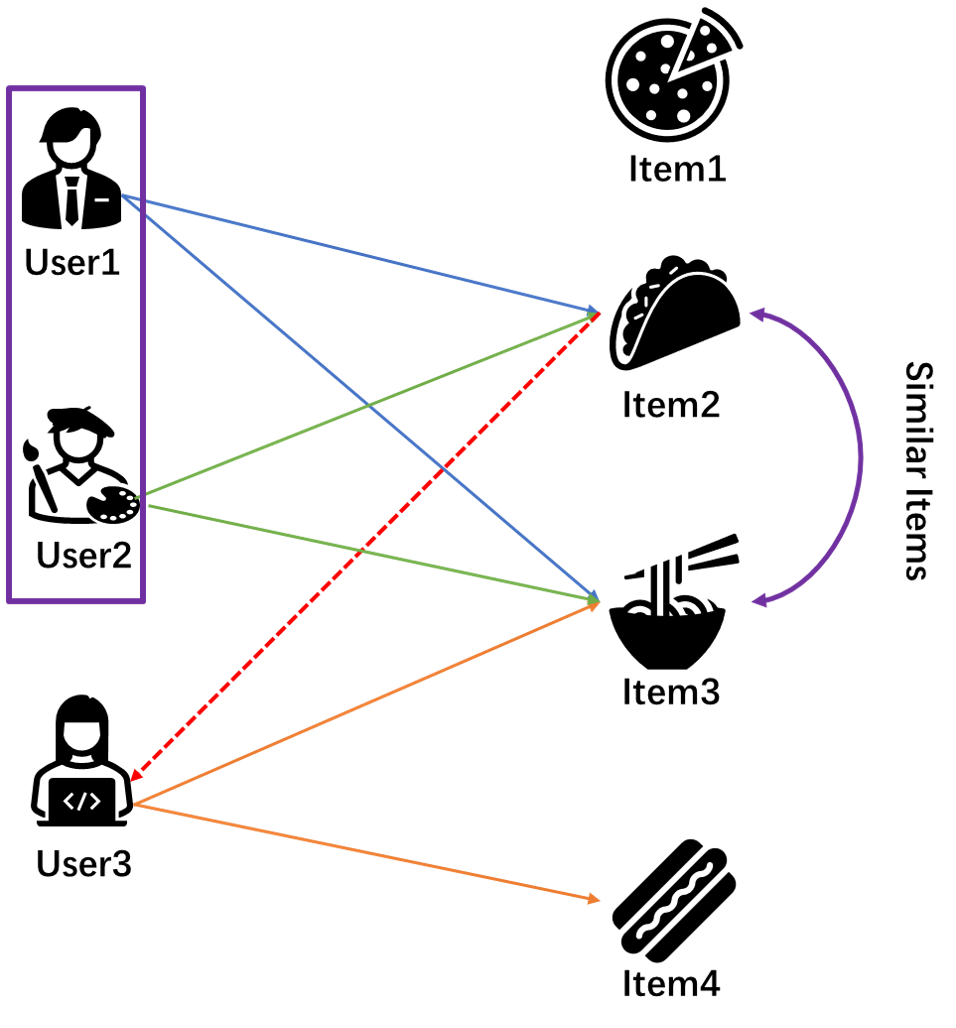}
\caption{Item-based collaborative filtering}
\end{figure}

We use Figure 2.3 to interpret the process of item-based collaborative filtering. The first step is to identify the similar item through the behavior of users instead of the contents of the items ( if we use the content of the item as the similarity criterion, then the system will morph into a content-based recommendation system). In the example in Figure 2.3, \textit{item2} and \textit{item3} are regarded as similar items due to this two items were received a positive rating from both \textit{user1} and \textit{user2}. By the figure, we can see that \textit{user3} shows the preference on \textit{item3}, so the item-based collaborative filtering algorithm will choose \textit{item2} which is the similar item of \textit{item3} to recommend to the \textit{user3}.
\subsection{Model-Based Methods}
It is worth noting that in the real world, the number of users is generally huge (e.g., Netflix has millions of users), compared to the number and type of items that are relatively fixed. In this scenario, calculating the similarity between items offline by using item-based collaborative filtering is preferred in order to optimize the efficiency of recommendation. On the contrary, in the case of a small number of users, user-based collaborative filtering is preferred. It can be seen from this that there is no good or bad recommendation system, and the important thing is to choose the right recommendation method in the right situation.

As can be seen from the above, memory-based methods are intuitive and highly interpretable models. Users can easily figure out why the memory-based system recommends these items to them (due to their similar users preferring this item or the item is similar to the one you have purchased or viewed). But on the other hand, the effect of memory-based methods to recommend popular products is more obvious. That is to say, if a lot of people like \textit{item1}, under memory-based methods, the probability of recommending it to you will be higher, although if item \textit{item2} also matches your interest, due to the rating for \textit{item2} is sparse, so the similarity of item \textit{item2} is greatly reduced when calculating the similarity.

So, is there a method that can handle sparse data with better generalization ability? Turning now to the methods to address the above question, which are called model-based methods.
\subsubsection{Latent Factor Models}
Model-based collaborative filtering has been studied for decades, and many methods have been proposed. Such as in \cite{multifacetedcf}, a multifaceted collaborative filtering model has been proposed or the probabilistic matrix factorization be described in \cite{Probabilistic}. In general, model-based collaborative filtering can produce a more accurate recommendation \cite{multifacetedcf}. In this section, matrix factorization (MF) which is the most popular latent factor model, will be illustrated.

The purpose of matrix factorization is to discover latent factor models of users and items in a shared latent space where the user-item relationship (e.g., user rating of the item) is calculated from the inner product of the user matrix and the item matrix.For a user-item rating matrix \textit{R} consisting of $m$ users and $n$ items, where $R\in\mathbb{R}^{m\times n}$. Through matrix factorization, the $m\times n$ user-item matrix \textit{R} be factorized into user latent matrix \textit{p} ($m\times k$ matrix) and item latent matrix \textit{q} ($k\times n$ matrix), where \textit{k} is the dimension of the \textit{latent factor}. In matrix factorization, the latent model of user \textit{u} and item \textit{i} are denoted by $p_{u}$ and $q_{i}$ respectively, where $p_{u} \in\mathbb{R}^{k}$, $q_{i} \in\mathbb{R}^{k}$. The notation ${r}_{u,i}$ is used to represent the rating of user \textit{u} on item \textit{i}. The predicted rating of user \textit{u} on item \textit{i} can be described as :
\begin{equation}
    \hat{r}_{u,i}=q_{i}^{T}p_{u}
\end{equation}
A general approach to training the latent model is to minimize the gap between predicted rating $\hat{r}_{u,i}$ and real rating $r_{u,i}$ that can be express as ${r}_{u,i}-q_{i}^{T}p_{u}$, by adopting \textit{root mean square error} (RMSE) as the loss function, we have:
    \begin{equation}
        \operatorname{Loss}=\min_{ q *, p * }\sum_{(u, i) \in K}\left(r_{u,i}-q_{i}^{T} p_{u}\right)^{2}
    \end{equation}
where $K$ denotes the set of $(u,i)$ pairs. By adopting $\mathit{l}_{2}$ regularization to void over-fitting problem, we have:
    \begin{equation}
        \operatorname{Loss}=\min_{ q *, p * }\sum_{(u, i) \in K}\left(r_{u, i}-q_{i}^{T} p_{u}\right)^{2}+\lambda\left(\left\|q_{i}\right\|+\left\|p_{u}\right\|\right)^{2}
    \end{equation}

\chapter{Methodologies}
In this chapter, we will discuss the techniques mentioned above in the manner that from the shallow to the deep. We will first discuss the methods individually, then we will diverge from a single method to a recommendation system that contains multiple methods, and finally, we will provide details of the proposed model, multi-armed bandit-based matrix factorization recommender system (BanditMF).
\section{Matrix Factorization}
As discussed above, to reduce the gap between predicted rating $\hat{r}_{u,i}$ and real rating $r_{u,i}$, \textit{root mean square error} (RMSE) is used as the loss function and $\mathit{l}_{2}$ norm is adopt as the regularization term. Then, when we already have the loss function, the next step is to minimize the loss function. Here we use the stochastic gradient descent (SGD) method to minimize the loss function. Next, the derivation of stochastic gradient descent will be provided. By Taylor expansions, we have the following equation:
\begin{equation}
    f(\theta) \approx f\left(\theta_{0}\right)+\left(\theta-\theta_{0}\right) \cdot f^{\prime}\left(\theta_{0}\right)
\end{equation}
where $\left(\theta-\theta_{0}\right)$ is a vector, it represents the distance of change. We can denote it by $\eta $. $\eta $ is a scalar, and the unit vector of $\left(\theta-\theta_{0}\right)$ is expressed by $v$:
\begin{equation}
    \left(\theta-\theta_{0}\right) = \eta \cdot v
\end{equation}
By formula 3.2, the formula 3.1 can change as:
\begin{equation}
        f(\theta) \approx f\left(\theta_{0}\right)+\eta v \cdot f^{\prime}\left(\theta_{0}\right)
\end{equation}
To minimize the loss function, we need minimize the $f(\theta)$. In other words, we want to make the value of the $f(\theta)$ smaller each time we update the $\theta$. So that $f(\theta)-f\left(\theta_{0}\right)< 0$, by this we have:
\begin{equation}
    f(\theta)-f\left(\theta_{0}\right) \approx \eta v \cdot f^{\prime}\left(\theta_{0}\right) < 0
\end{equation}
Where, $\eta $ is scalar and generally positive, which can be ignored. Therefore:
\begin{equation}
    v \cdot f^{\prime}\left(\theta_{0}\right) < 0
\end{equation}
From vector multiplication, when the angle between two vectors is 180 degrees, the product of vector multiplication is less than 0, and minimal. So, when the direction of $v$ and $f^{\prime}\left(\theta_{0}\right)$ is opposite (i.e. included angle is 180 degrees), $v \cdot f^{\prime}\left(\theta_{0}\right) < 0$ and have minimal value. From this we can get: 
\begin{equation}
    v=-\frac{f^{\prime}(\theta)}{\left\|f^{\prime}(\theta)\right\|}
\end{equation}
where $v$ is a unit vector, so we divide it by the norm of the $f^{\prime}(\theta)$, substitute the formula 3.6 into formula 3.2:
\begin{equation}
    \theta=\theta_{0}-\eta \frac{f^{\prime}(\theta)}{\left\|f^{\prime}(\theta)\right\|}
\end{equation}
Due to the norm of a vector being a scalar, we obtain:
\begin{equation}
    \theta=\theta_{0}-\eta f^{\prime}(\theta)
\end{equation}
where $\theta_{0}$ is is the current parameter, $\eta$ denote the stride or so call learn rate, $f^{\prime}(\theta)$ is the direction of the gradient.
However, our loss function is a multivariate function, so we need to calculate the partial derivatives of it. So the formula of $\theta$ become:
\begin{equation}
    \theta=\theta_{0}-\eta  \frac{\partial}{\partial \theta_{0}} J(\theta)
\end{equation}
For our loss function,
\begin{equation}
        \text { Loss }: J(p, q)=\min_{ q *, p * } \sum_{(u, i) \in K}\left(r_{u,i}-q_{i}^{T} p_{u}\right)^{2}+\lambda\left(\left\|q_{i}\right\|+\left\|p_{u}\right\|\right)^{2}
\end{equation}
Where $q_{i}$ and $p_{u}$ are the variables which need to optimize, $\lambda$ is the regularization factor. Our goal is to optimize the $q_{i}$ and $p_{u}$ and find the minimum value of the loss function.
Perform partial derivative on both $q_{i}$ and $p_{u}$:
    \begin{equation}
        \frac{\partial E}{\partial q_{i}}=-2\left(r_{u,i}-q_{i}^{T} p_{u}\right) p_{u}+2 \lambda q_{i}
    \end{equation}
    \begin{equation}
        \frac{\partial E}{\partial p_{u}}=-2\left(r_{u,i}-q_{i}^{T} p_{u}\right) q_{i}+2 \lambda p_{u}
\end{equation}
Substituted two partial derivatives into the gradient descent formula 3.9:
\begin{equation}
    q_{i}=q_{i}+2 \eta\left(\left(r_{u,i}-q_{i}^{T} p_{u}\right) p_{u}-\lambda q_{i}\right)
\end{equation}
\begin{equation}
    p_{u}=p_{u}+2 \eta\left(\left(r_{u,i}-q_{i}^{T} p_{u}\right) q_{i}-\lambda p_{u}\right)
\end{equation}
By define $e_{u,i}=r_{u,i}-q_{i}^{T} p_{u}$. we have:
\begin{equation}
    q_{i} \leftarrow q_{i}+\eta\left(e_{u,i} p_{u}-\lambda q_{i}\right)
\end{equation}
\begin{equation}
    p_{u} \leftarrow p_{u}+\eta\left(e_{u,i} q_{i}-\lambda p_{u}\right)
\end{equation}
To sum up, the above process is the derivation process of stochastic gradient descent.By the stochastic gradient descent algorithm, we train two latent vector matrices: user matrix \textit{p} and item matrix \textit{q}. Through the inner product of user matrix \textit{p} and item matrix \textit{q}, we can have the predicted user-item matrix. If the user matrix and item matrix are fitted by the loss function perfectly, then our predicted rating will be infinitely close to the real rating. 

In this project, we consider another scenario in which some harsh users give a low rating, but some users are more tolerant, giving a relevant high score for bad items. By this intuition, we trying to add bias into matrix factorization.
\begin{equation}
        \hat{r}_{u,i}=\underset{basic}{\underbrace{q_{i}^{T}p_{u}}}+ \underset{bias}{\underbrace{b_{ui}}}
    \end{equation}
So that our predicted rating become:
\begin{equation}
    \hat{\boldsymbol{r}}_{u,i}=b_{u,i}+q_{i}^{T} * p_{u}
\end{equation}
And the loss function is:
\begin{equation}
    \operatorname{Loss}=\min _{q^{*}, p^{*}} \sum_{(u, i) \in K}\left(r_{u,i}-b_{u,i}-q_{i}^{T}p_{u}\right)^{2}+\lambda \left(\left\|q_{i}\right\|^{2}+\left\|p_{u}\right\|^{2}+b_{u}^{2}+b_{i}^{2}\right)
\end{equation}
where $b_{u,i} = \mu+b_{u}+b_{i}$. We use $\mu$ to denote the average rating for the entire rating matrix, $b_{u}$ denote the user's bias, and $b_{i}$ is represent the item bias.
In the next chapter, we will simulate these two approaches with real data and evaluate the performance between the two approaches .
\section{Memory-Based Collaborative Filtering Recommender System}
As was pointed out in the formulation of this report, memory-based collaborative filtering makes recommendations based on the similarity between the users or the items. The detailed steps of the memory-based collaborative filtering can be stated as follows:
\begin{itemize}
    \item[1. ]Identify a target user who will be recommended.
    \item[2. ]Figure out the items that the target user has given ratings for.
    \item[3. ]Based on the target user’s rating pattern, find similar users.
    \item[4. ]Get the rating records of similar users for items.
    \item[5. ]Calculate the similarity between the target user and a similar user by formula (e.g., \textit {Euclidean Distance}, \textit{Pearson Correlation}, \textit{Cosine Similarity}).
    \item[6. ]Recommend the items with the highest score to the target user.

\end{itemize}
In the following, an example will be demonstrated to elaborate on the above steps.
\begin{figure}[htbp]
\centering
\includegraphics[scale=0.55]{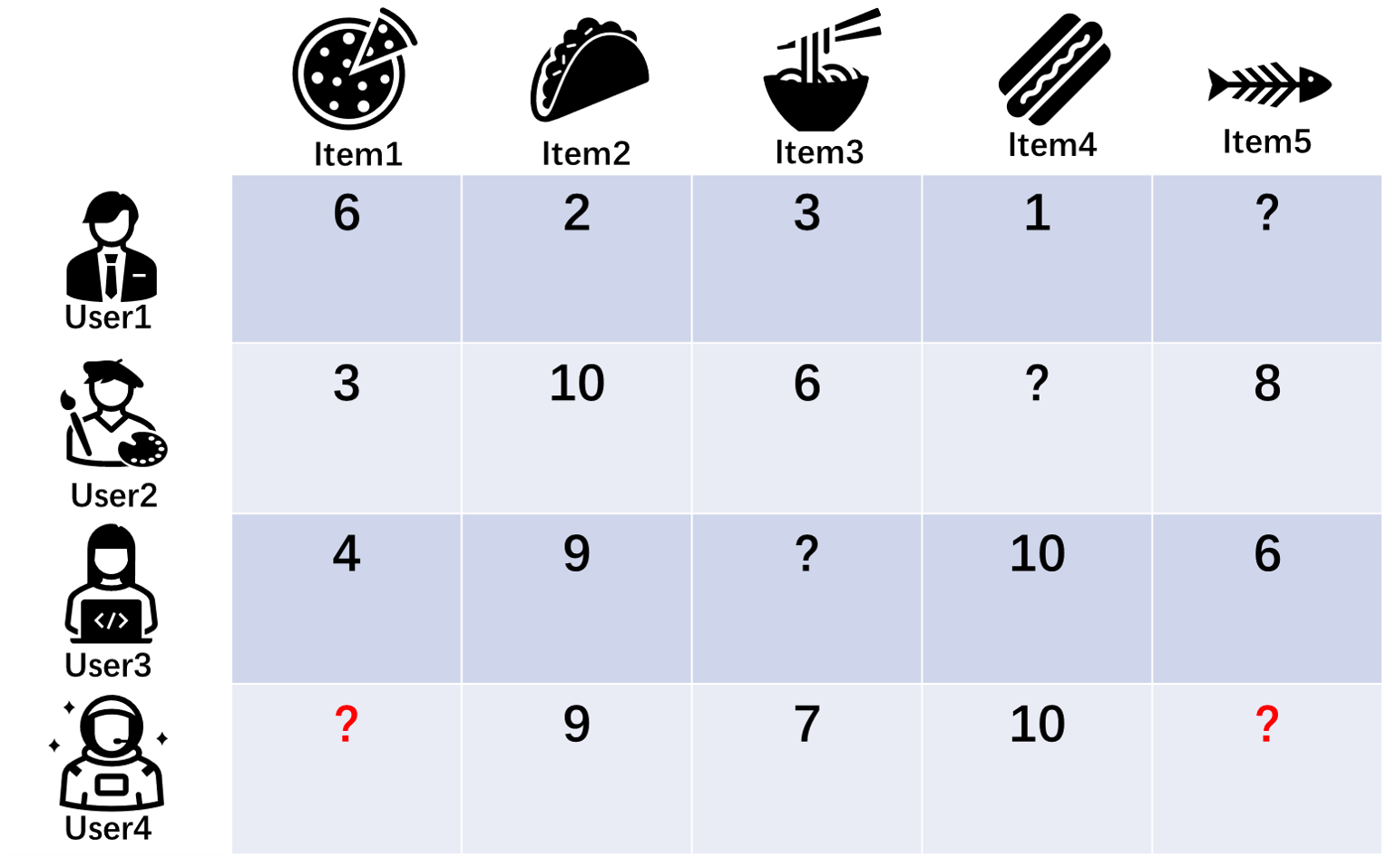}
\caption{Example of User$-$item rating matrix}
\end{figure}

As per the user-item rating matrix shown in Figure 3.1, assume \textit{user4} is our target user and rated three out of five items. The goal of the system is to figure out which one of the two unrated items (i.e., \textit{item1} and \textit{item5}) should be recommended to the target user who can probably get a high rating. After we define our target user and figure out some users, the next step that comes up is to calculate the similarity between the target user and similar users. The common methods of similarity calculation include \textit{Euclidean Distance}, \textit{Pearson Correlation}, \textit{Cosine Similarity}. To calculate the similarity between two users, we base it on the ratings of \textit{item2}, \textit{item3}, and \textit{item4} which are the items that all users have rated.
\begin{figure}[htbp]
\centering
\includegraphics[scale=0.5]{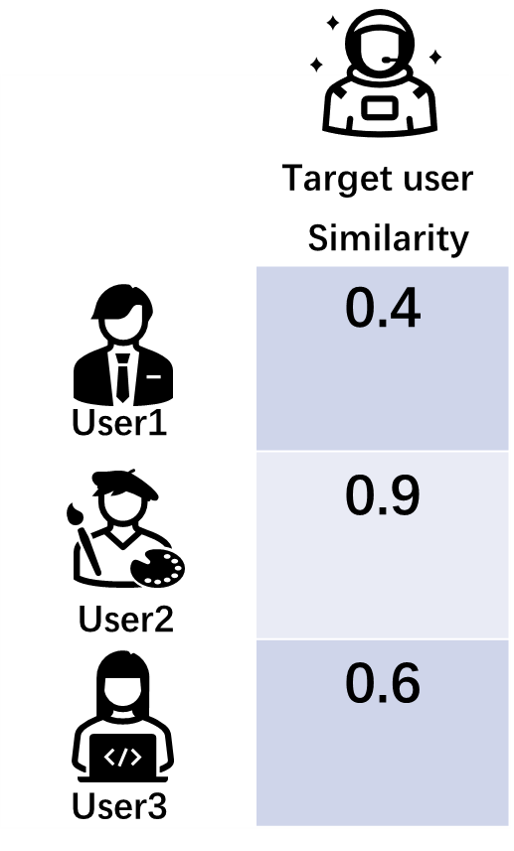}
\caption{Similarity between target user and other users}
\end{figure}
Here we make a hypothesis that the similarity between the target user and other users is 0.4, 0.9, and 0.6, respectively, as shown in Figure 3.2. The next step is that, based on the value of similarity, we can calculate the predicted rating for the target user. What can be seen in Figure 3.3 is the so-called weighted rating matrix because it provides more weight to those users who have a higher similarity to the target user.
\begin{figure}[htbp]
\centering
\includegraphics[scale=0.5]{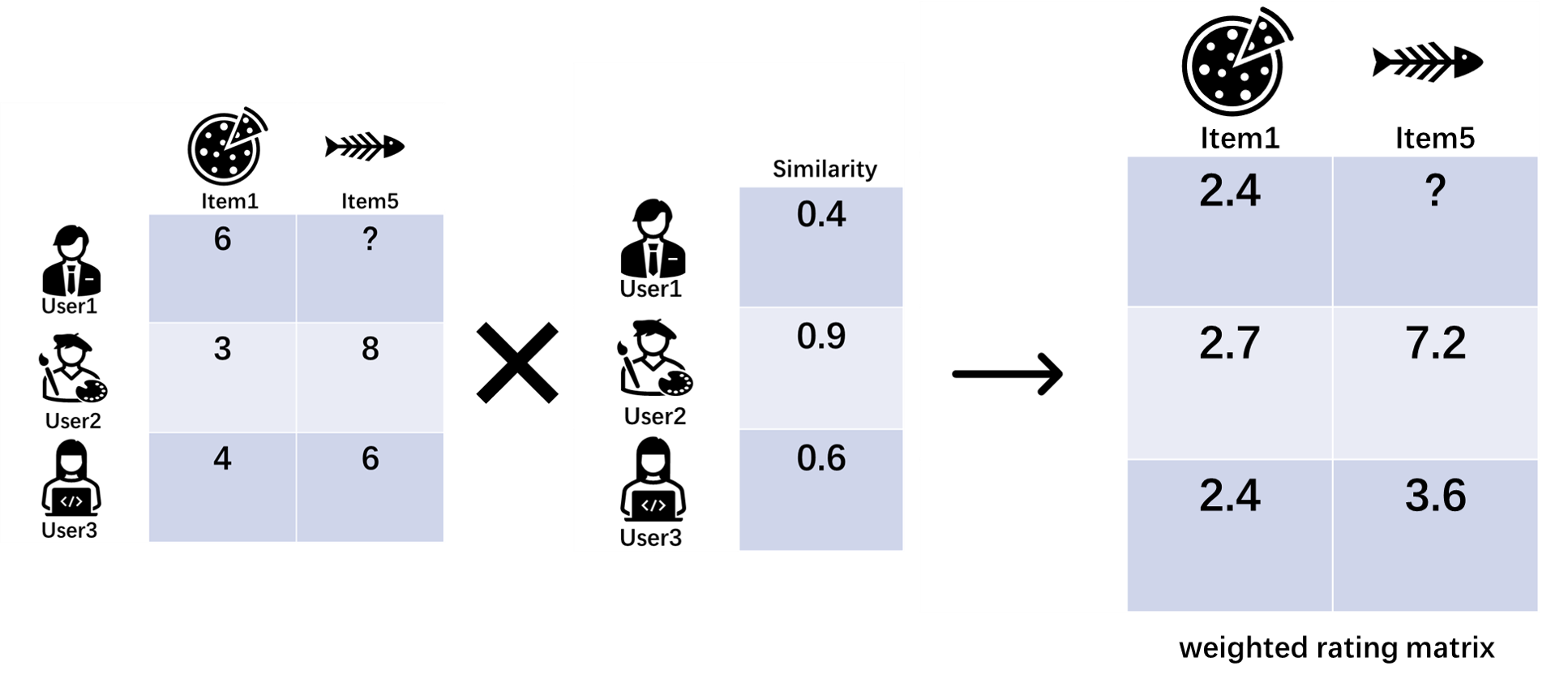}
\caption{Weighted rating matrix}
\end{figure}
For the final step, we tally up all weighted rates to create the recommender matrix. Due to the sparsity of the weighted rating matrix (e.g. only \textit{user2} and \textit{user3} provided the rating for \textit{item5} , we normalize the value of the weighted rating by dividing the sum of weighted ratings by the sum of the similarity for users.
\begin{figure}[htbp]
\centering
\includegraphics[scale=0.5]{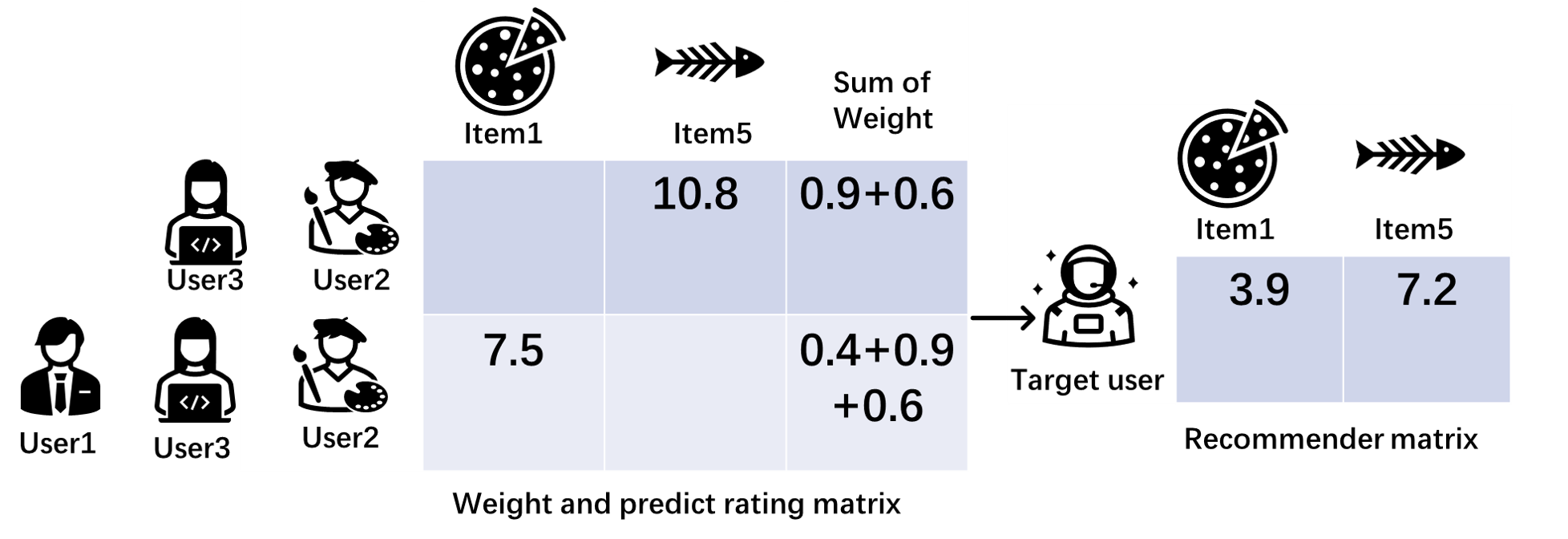}
\caption{Recommender matrix}
\end{figure}
According to the result of user-based collaborative filtering shown in Figure 3.4, the value in the recommender matrix describes the potential rating for the target user that will arise for the items based on the similarity between them and other users. The target user potentially provides a higher rating to \textit{item5} than \textit{item1}. So, through the result, the user-based collaborative filtering recommender system will choose \textit{item5} to recommend to the target user.

\section{Hybrid Recommender System: Combination of Content-Based Method and Matrix Factorization}
In the above paper, matrix factorization has been discussed. For the content-based method, the concept is straightforward, which is to recommend items that are similar to what the target user liked before. Now, the intuition is whether we can combine the two approaches into a hybrid recommender system. Recall that in matrix factorization, by the stochastic gradient descent algorithm, we train two latent vector matrices: user matrix \textit{p} and item matrix \textit{q}. Through the inner product of user matrix \textit{p} and item matrix \textit{q}, we can have the predicted user-item matrix. In the predicted user-item matrix, row vectors represent the preferences of each user, and column vectors represent the attributes of each item, which means the user similarity can be obtained by calculating the similarity between row vectors, and similarly, the item similarity can be obtained by calculating the similarity of column vectors.

For the proposed hybrid system, the content-based approach is performed based on the matrix factorization (i.e., the content-based approach calculates the similarity through the user-item rating matrix derived by the result of matrix factorization). Consider that the detailed methodology of matrix factorization is already stated in section 3.1, so there we will illustrate the methodology after obtaining the user-item rating matrix, which derives from the result of matrix factorization.

Assume the predicted user-item rating matrix is available to recommend items to the target user. We have the following steps:
\begin{itemize}
    \item[1. ]  List all the item indexes rated by the target user.
    \item[2. ] Sort the items that have not yet been rated by the target user (i.e., the item indexes are not listed in the \textit{step1}) through our predicted rating score.
    \item[3. ] Get the top ranked item of the list created by \textit{step2}.
\end{itemize}
Then we use the \textit{cosine similarity} for similarity measurement to measure how similar the items are in attractiveness to the target user, where:
\begin{equation}
\operatorname{similar}(i, j)=\cos (i, j)=\frac{i \cdot j}{\|i\| \cdot\|j\|}
\end{equation}
The formula uses the angle between two vectors to calculate the cosine value. The angle between the vectors $i$ and $j$ is the thing this formula cares about, instead of the magnitude. So that the value of similarity is within the interval [-1,1], and 1 means “exactly the same”. After calculating the cosine similarity, we sort the items by the value of similarity, and then we can get the items that are most similar to the target user's favorite items. The implementation of this hybrid recommender system with real-world data will be presented in the next chapter.

It is foreseeable that if it combines memory-based collaborative filtering with a content-based approach, it can solve the problems faced by memory-based collaborative filtering, such as sparsity and loss of information. Similarly, as can be seen from our method description, hybrid systems increase computational steps and complexity.
\section{BanditMF: Multi-Armed Bandit Based Matrix Factorization Recommender System}
The authors in \cite{billsus2000user} proposed a method that is based on implicit and explicit user feedback, agents use machine learning algorithms to inscribe an individual user model and use the user model to make recommendations. The method called BanditMF, a matrix factorization recommendation system based on multi-arm bandit, is proposed to address the limitations of collaborative filtering and bandit algorithms as mentioned earlier. The system combines model-based collaborative filtering matrix factorization (MF) with a multi-armed bandit algorithm. BanditMF incorporates an offline subsystem centered on matrix factorization and an online subsystem focused on a multi-armed bandit algorithm.

\begin{minipage}{\textwidth}
        \begin{minipage}[t]{0.45\textwidth}
            \centering
            \makeatletter\def\@captype{table}\makeatother
            \begin{tabular}{llllll}
\hline
                       & $i_{1}$ & $i_{2}$   & $i_{5}$   & $i_{4}$   & $i_{5}$   \\ \hline
\multicolumn{1}{l}{$u_{1}$} & 1.4      & ?   & 1.1 & 0.7 & ?   \\
\multicolumn{1}{l}{$u_{2}$} & ?        & 0.3 & ?   & 0.7 & 0.5 \\
\multicolumn{1}{l}{$u_{3}$} & 0.4      & 0.3 & ?   & ?   & 0.3 \\
$u_{4}$                     & 1.4      & ?   & 1.2 & ?   & 0.8 \\ \hline
\end{tabular}
\caption{User-item rating matrix}
        \end{minipage}
        \begin{minipage}[t]{0.45\textwidth}
        \centering
        \makeatletter\def\@captype{table}\makeatother
\begin{tabular}{llllll}
\hline
                       & $i_{1}$ & $i_{2}$   & $i_{5}$   & $i_{4}$   & $i_{5}$   \\ \hline
\multicolumn{1}{l}{$u_{1}$} & 1.4      & 0.8   & 1.1 & 0.7 & 0.9   \\
\multicolumn{1}{l}{$u_{2}$} & 1.0       & 0.3 & 1.0   & 0.7 & 0.5 \\
\multicolumn{1}{l}{$u_{3}$} & 0.4      & 0.3 & 0.3   & 0.1   & 0.3 \\
$u_{4}$                     & 1.4      & 0.7   & 1.2 & 0.8   & 0.8 \\ \hline
\end{tabular}
\caption{Predicted rating matrix}
        \end{minipage}
    \end{minipage}

By analyzing the user-item rating matrix as shown in Table 3.1, we can obtain the predicted user-item rating matrix as shown in Table 3.2.
For the sake of explanation, we consider each row in Table 3.2 as a predictive model. In another world, we consider the row vectors that represent the preferences of each user as models. When a new user enters the system, a predictive model is selected and used as a basis to make recommendations for the new user. However, in the real world, the number of users is generally in the millions, so the above approach is rendered impractical due to the $m\times n$ scaling rating matrix, where the $m$ is millions of orders in size. 

\begin{figure}[htbp]
\centering

\includegraphics[scale=0.8]{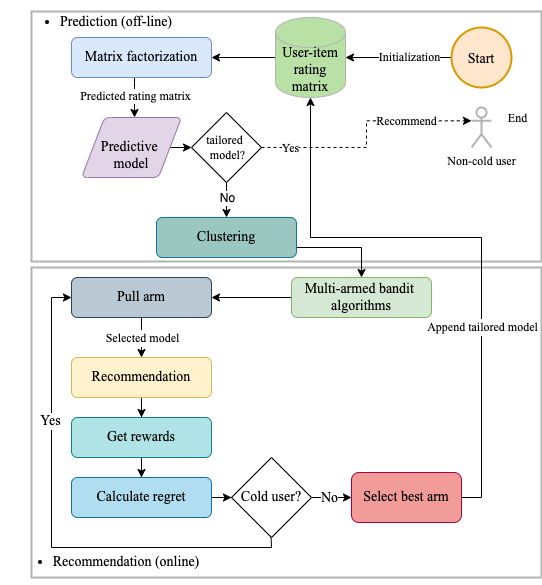}
\caption{Flow chart of the BanditMF}
\end{figure}

Whether we can cluster the rows in the matrix to reduce the size of the rating matrix(i.e., reduce the number of predictive models)? Inspired by collaborative filtering, it is possible to cluster the rows of the rating matrix such that each row of the matrix represents a row vector, and each row vector represents the preferences of each user, so that each cluster is composed of users with similar rating preferences. Another important reason for matrix factorization is the very widespread assumption that the user-item rating matrix is a low-rank matrix, so the clustering can be performed on the rows of the user-item rating matrix. We treat BanditMF as an algorithm consisting of two stages: (1) prediction, and (2) recommendation.

For the offline subsystem, we define the prediction model as the following steps: matrix factorization based rating prediction, clustering, and unified rating

\textbf{\textit{Matrix factorization based rating prediction}}. As we illustrated in section 3.1, the non-cold user formed the user-item rating matrix $ R $and, based on the matrix factorization, we can get the predicted user-item rating matrix $\hat{R}$. The notation $\hat{r}_{u,i}$ is used to represent the rating of user \textit{u} on item \textit{i}. Recall that in this project, our predicted rating is:
\begin{equation}
    \hat{\boldsymbol{r}}_{u,i}=b_{u,i}+q_{i}^{T} * p_{u}
\end{equation}
And the loss function is:
\begin{equation}
    \operatorname{Loss}=\min _{q^{*}, p^{*}} \sum_{(u, i) \in K}\left(r_{u,i}-b_{u,i}-q_{i}^{T}p_{u}\right)^{2}+\lambda \left(\left\|q_{i}\right\|^{2}+\left\|p_{u}\right\|^{2}+b_{u}^{2}+b_{i}^{2}\right)
\end{equation}
where $b_{u,i} = \mu+b_{u}+b_{i}$. We use $\mu$ denote the average rating for the entire rating matrix, $b_{u}$ denote the user's bias, and $b_{i}$ is represent the item bias.

\textbf{\textit{Clustering}}. According to the user’s preference vectors of the predicted user-item rating matrix $\hat{R}$, we can perform the clustering on the row of the $\hat{R}$.  In this system, \textit{K-means} algorithm is adopted, which is measured by \textit{Euclidean distance}. After performing the clustering on the user’s preference vectors, we will have a set of clusters where $\mathcal{C}=\left \{ c_{1},c_{2}\cdots c_{n} \right \}$.

\textbf{\textit{Unified rating}}. For the clusters in $\mathcal{C}$, we perform the average rating on it to get the unified user preference vectors, where $\hat{\alpha}$ denote the unified user preference vectors. For example, as shown in Table 3.3, by the hypothetical that $u_{1}$ and $u_{2}$ be clustered into $ c_{1}$, the unified user preference vectors $ \hat{\alpha}=\left \{ 1.2,0.55, 1.05,0.7, 0.7 \right \}$, which is the average of the $u_{1},u_{2}$.

\begin{table}[htbp]
    \centering
\begin{tabular}{llllll}
\hline
  & $i_{1}$ & $i_{2}$   & $i_{5}$   & $i_{4}$   & $i_{5}$   \\ \hline
$u_{1}$& 1.4      & 0.8   & 1.1 & 0.7 & 0.9   \\
$u_{1}$& 1.0       & 0.3 & 1.0   & 0.7 & 0.5 \\ \hline
$\hat{\alpha}_{1}$& 1.2      & 0.55 & 1.05   & 0.7   & 0.7 \\

\end{tabular}
\caption{Unified user preference vectors}
\end{table}
The online recommendation module is mainly composed of multi-armed bandits, and this online module runs as follows:
\begin{itemize}
    \item[1. ] Select a predictive model $\mathcal{S}_{n}$ using the MAB algorithm $\pi$
    \item[2. ]  Select and recommend the item $i^*$ with the top rating score in $\hat{\alpha}$ which is not recommended to user $u$.
    \item[3. ]  Receive the user $u$'s rating for item $i^*$.
    \item[4. ]  calculate the reward and regret.
    \item[5. ]  Determine whether a user is a cold user by a threshold $\tau $.
    \item[6. ]  If yes, update the algorithm $\pi$. Otherwise, turn to the offline module.
\end{itemize}
The online recommendation module will provide recommendations to new users until the new user is no longer cold (i.e., some feedback on the recommended items or preferences has been provided to the system). Nevertheless, if a system obliges the user to make excessive ratings, the user will lose interest in the system \cite{threshold}. For the above reasons, we cannot run the bandit algorithm too many times to get the optimal recommendation, and online systems need to get user preferences in a limited number of recommendations(i.e., within the threshold $\tau $).

Now let's walk through the entire BanditFM workflow from scratch, which is the flowchart shown in Figure 3.5. The input information is first fed into the offline prediction module, and through matrix factorization, we get the predicted user-item rating matrix. Then each row of the matrix, namely the user preference vector, is clustered to reduce the size of the prediction matrix. If we denote the original size predicted user-item rating matrix as $m\times n$, after clustering, we have a ${m}'\times n$ matrix, where ${m}'\ll m$. Then pass the clustered predictive model to the online recommendation module. Through the multi-armed bandit algorithm, it selects the predictive model from the clusters. Get user preferences through interaction with users (i.e., an algorithm recommends items to users and users provide feedback). The online module determines whether a user is a cold user through a threshold $\tau $, and if the system has collected certain user rating preferences, then the user is no longer a cold user. At this point, we switch to the offline prediction module and append the user preferences obtained from the online module to the user-item rating matrix. We can then accurately make recommendations to new users through the matrix factorization technique.

\textbf{\textit{Reward and Regret}}. In BanditMF, we define the \textit{reward} as the equation shown in the formula 3.23.
\begin{equation}
    \mu_{\mathcal{S}_{n}}\left ( t \right ) =\frac{r_{u,i}}{r^*}
\end{equation}
where $r^*$ denote the maximum rating in the dataset. $r_{u,i}$ denote the rating of user $u$ for recommended item $i$ according to the algorithm selected predictive model $\mathcal{S}_{n}$. The reward interval is $\mu_{\mathcal{S}_{n}}\left ( t \right )\in [0,1]$.This intuition of this reward can be interpreted as the higher the feedback received for the recommendation, the higher the reward will be. If by the reward, then we define the \textit{regret} as:
\begin{equation}
    R(T) \triangleq \sum_{t=1}^{T}\left(\mathbb{E}\left[\mu^*\right]-\mathbb{E}\left[\mu_{\mathcal{S}_{n}}\left ( t \right )\right]\right)
\end{equation}
where $\mathbb{E}\left[\mu^*\right]$ denotes the maximum expected cumulative rewards.

In the next chapter, we will verify the BanditMF using various criteria and compare the performance of different multi-armed bandit algorithms.

\chapter{Experiments and Implementation}
In this chapter, the methods mentioned in the previous chapters will be implemented and validated with different real-world datasets.
\section{Contextual Bandit: LinUCB}
We introduced the context-free algorithm in detail last semester, so we will not continue to elaborate on it in this report. Instead, we will focus on the contextual bandit, implementing and verifying one of the most famous algorithms – LinUCB which was proposed by the authors in \cite{context}. LinUCB is an extension of the UCB algorithm. The algorithm chooses an arm based on observation of the current user and set of arms together with the context feature.  Also, in the LinUCB algorithm, there exists a factor $\alpha$, which is used to control the trade-off between exploration and exploitation.

\begin{figure}[htbp]
\centering
\includegraphics[scale=0.8]{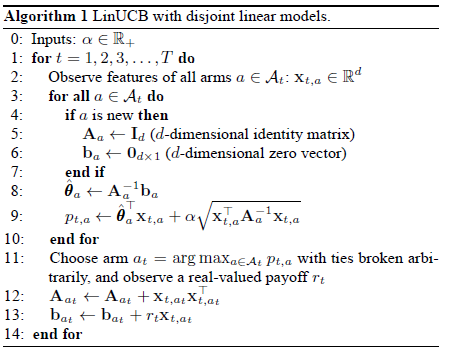}
\caption{LinUCB with disjoint linear models \cite{context}}
\end{figure}

\begin{figure}[htbp]
    \centering
    \includegraphics[scale =0.4]{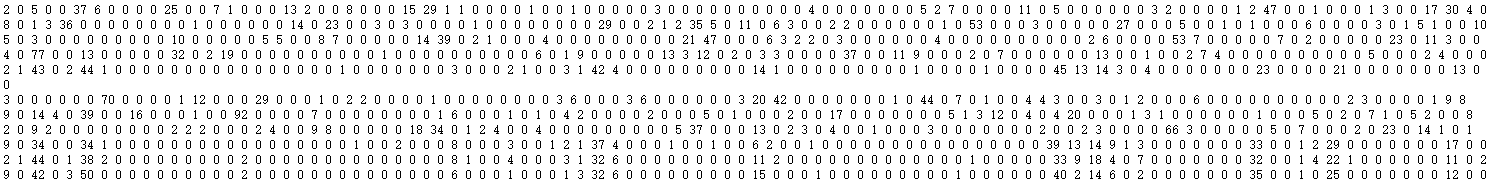}
    \caption{Format of the personalization dataset}
    \end{figure}
\textbf{\textit{Dataset}}. For this experiment, the personalization dataset\footnote[3]{Dataset from: http://www.cs.columbia.edu/~jebara/6998/dataset.txt} has been used. Figure 4.2 demonstrates the format of the dataset. This dataset consists of a matrix of size $10000 \times 102$. The data in the first column indicates the action has been chosen. The second column is value indicate the obtained reward, where $y_{t}$ denote the real reward that was obtained at time $t$ and $y_{t}\in \mathbb{B}$. From column 3 to column 103 are the context which is represented as the feature vector $\mathbf{x_{t}}\in\mathbb{R}^{100}$.

\textbf{\textit{Compute click-through rate (CTR)}}. The click-through rate (CTR) proposed in \cite{ctr} is used to evaluate the performance of the algorithm. The click-through rate replay at time $T$ is:
    \begin{equation}
        C(T)=\frac{\sum_{t=1}^{T} y_{t} \times \mathbf{1}\left[\pi_{t-1}\left(\mathbf{x}_{t}\right)=a_{t}\right]}{\sum_{t=1}^{T} \mathbf{1}\left[\pi_{t-1}\left(\mathbf{x}_{t}\right)=a_{t}\right]}
    \end{equation}
where, $\pi_{t-1}$ denote the algorithm trained on data up to time $t-1$. At time $t$, if the context $\mathbf{x}_{t}$ been observed by algorithm $\pi_{t-1}$, then we use the notation $\pi_{t-1}(\mathbf{x}_{t})$ denote the chosen action. $a_{t}$ and $y_{t}$ denote the real action that was taken in the dataset at time $t$ and the real reward that was obtained at time $t$ respectively.

\textbf{\textit{Experiment}}. In this experiment (ideas and parameters are referenced from \cite{experiment}), the size of the context is 100, where $\mathbf{x_{t}}\in \mathbb{R}^{100}$, the number of arms $K=10$, total iteration $T=10000$. For the $\alpha$, as the only input in the algorithm, optimizing this parameter may result in higher total payoffs in practice. So, we will experiment with the blow values of $\alpha$:
\begin{enumerate}
   
    \item For Alpha = 1
     \item For Alpha = 0.25
    \item For Alpha = ${1}/{\sqrt{t}}$
    \item For Alpha = 0.001
    \item For Alpha = $0.001/(0.1 * Prediction Count)$
    
\end{enumerate}
\begin{figure}[htbp]
    \centering
    \includegraphics[scale =0.7]{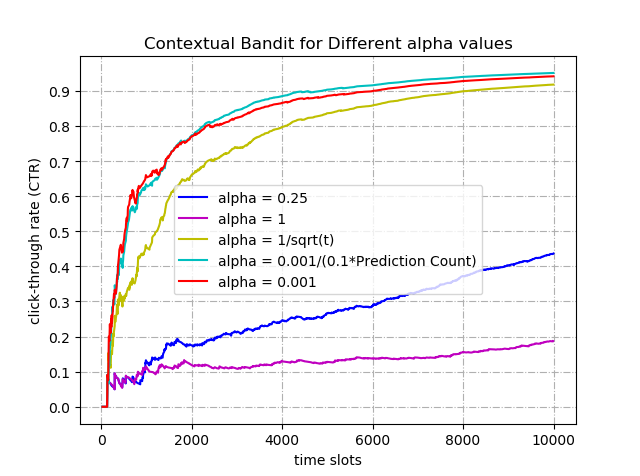}
    \caption{CTR of LinUCB for varying Alphas}
    \end{figure}
The click-through rate (CTR) proposed in A is used to evaluate the performance of the algorithm. As it is evident from Figure 4.3, the best CTR value was achieved with an alpha equal to $0.001/(0.1 * Prediction Count)$ when the CTR value was 0.95, where the prediction count is the number of correct predictions.

To understand the trade-off between exploration and exploitation, for each alpha value, first the mean value of the upper confidence bound for 10 arms will be plotted. We will then compare the number of predictions made by each arm with the number of true predictions it made.
\begin{frame}{}
\begin{figure}[htbp]
\centering
\subfigure[MeanUCB]{
\begin{minipage}[t]{0.5\linewidth}
\centering
\includegraphics[scale=0.5]{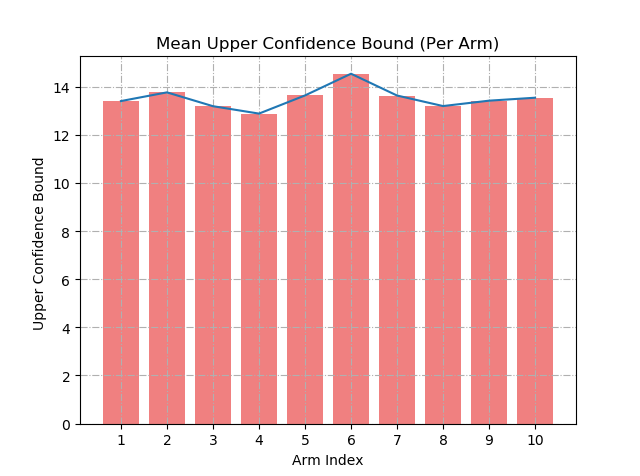}
\end{minipage}%
}%
\subfigure[Prediction comparison]{
\begin{minipage}[t]{0.5\linewidth}
\centering
\includegraphics[scale=0.5]{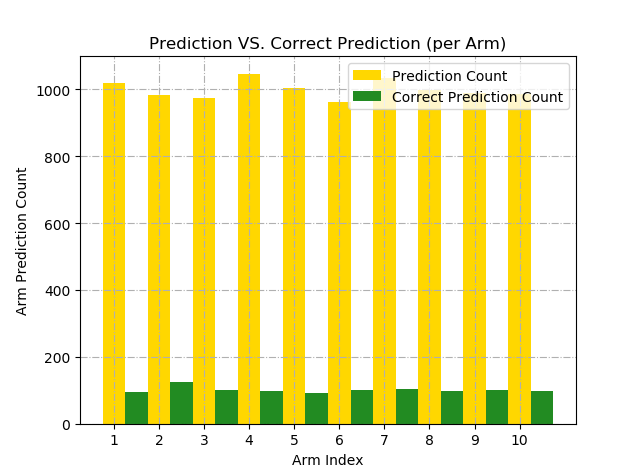}
\end{minipage}%
}%
\centering
\caption{ Mean UCB and Total VS. Correct prediction for
Alpha = 1}
\end{figure}  
\end{frame}

As it is evident from Figure 4.4, when the $\alpha$ value is 1, the exploration and exploitation here is at a minimum, which can be represented by strips of nearly the same length. That implies that the upper confidence bound of each arm is very approximate. For this cause, we can also observe that the predicted counts are nearly the same for all arms,thus when the $\alpha$ value equals to 1, the algorithm giving the lowest click rate value which is 0.18 as shown in Figure 4.3.
\begin{figure}[htbp]
\centering
\subfigure[MeanUCB]{
\begin{minipage}[t]{0.5\linewidth}
\centering
\includegraphics[scale=0.5]{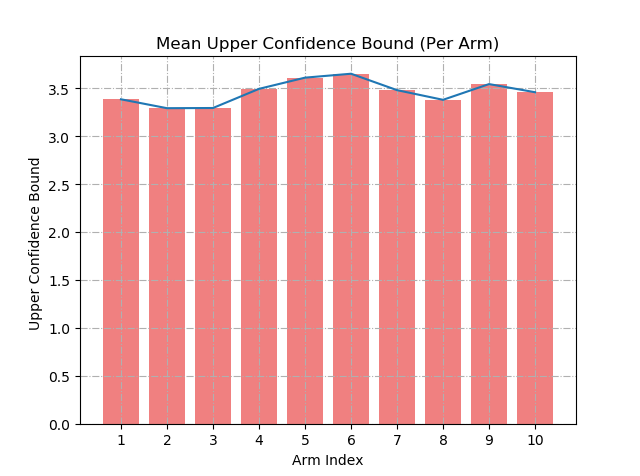}
\end{minipage}%
}%
\subfigure[Prediction comparison]{
\begin{minipage}[t]{0.5\linewidth}
\centering
\includegraphics[scale=0.5]{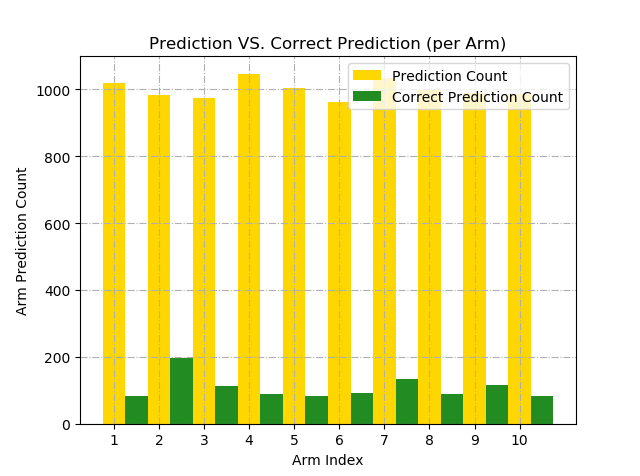}
\end{minipage}%
}%
\centering
\caption{ Mean UCB and Total VS. Correct prediction for
Alpha = 0.25}
\end{figure}  
For Figure 4.5, When the $\alpha$ value change to 0.25, the improvement is not significant compare to the result when $\alpha$ value equals to 1 .
\begin{figure}[htbp]
\centering
\subfigure[MeanUCB]{
\begin{minipage}[t]{0.5\linewidth}
\centering
\includegraphics[scale=0.5]{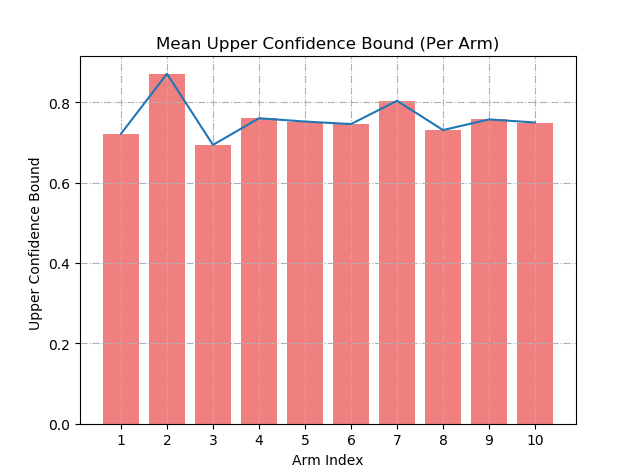}
\end{minipage}%
}%
\subfigure[Prediction comparison]{
\begin{minipage}[t]{0.5\linewidth}
\centering
\includegraphics[scale=0.5]{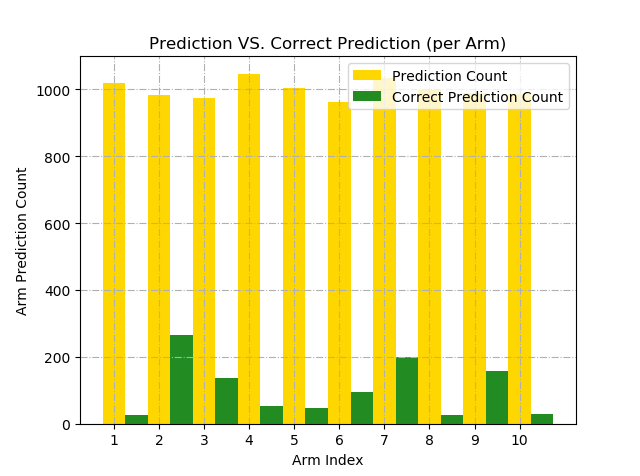}
\end{minipage}%
}%
\centering
\caption{ Mean UCB and Total VS. Correct prediction for
Alpha = ${1}/{\sqrt{t}}$}
\end{figure}  
Now, when we change our $\alpha$ to the time-dependent values (i.e., square root of the time), as shown in Figure 4.6, the result receives a significant improvement. This is mainly caused by the fact that we adjust our exploration according to the changes in time and limit our predictions in order to maximize the use of our trained algorithms over time.
\begin{figure}[htbp]
\centering
\subfigure[MeanUCB]{
\begin{minipage}[t]{0.5\linewidth}
\centering
\includegraphics[scale=0.5]{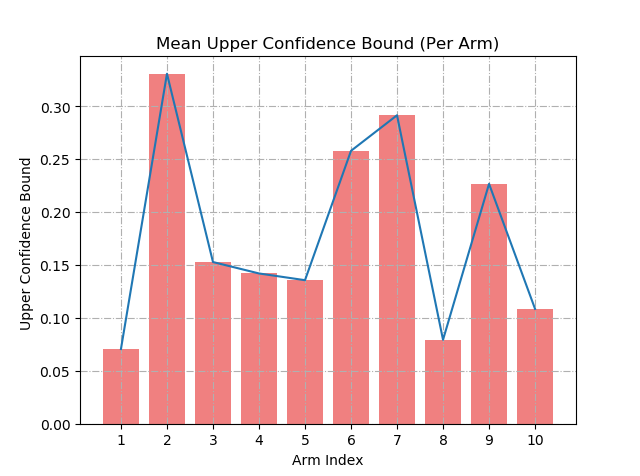}
\end{minipage}%
}%
\subfigure[correct prediction comparison]{
\begin{minipage}[t]{0.5\linewidth}
\centering
\includegraphics[scale=0.5]{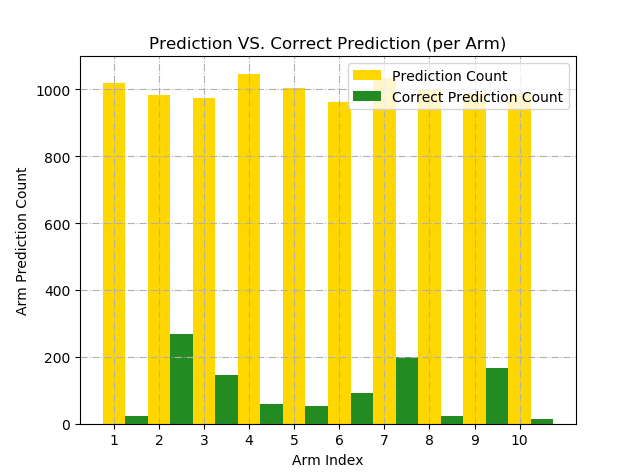}
\end{minipage}%
}%
\centering
\caption{ Mean UCB and Total VS. Correct prediction for
Alpha = 0.001}
\end{figure}  
Figure 4.7 shows that when $\alpha=0.001$, a higher CTR can get. Because of the extremely small value of $\alpha$, the algorithm will perform exploiting most of the iteration and limiting the exploration.
\begin{figure}[htbp]
\centering
\subfigure[MeanUCB]{
\begin{minipage}[t]{0.5\linewidth}
\centering
\includegraphics[scale=0.5]{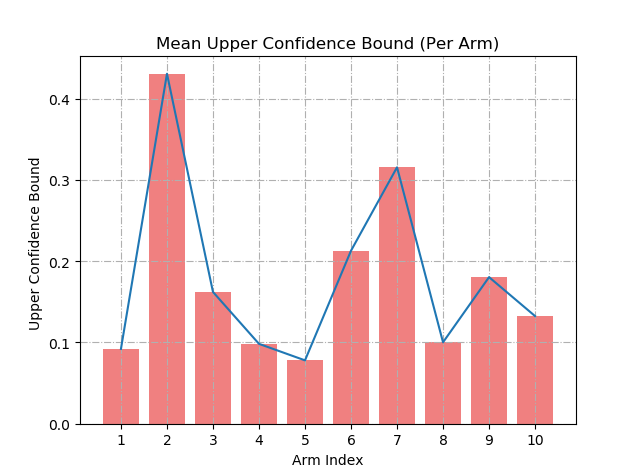}
\end{minipage}%
}%
\subfigure[Prediction comparison]{
\begin{minipage}[t]{0.5\linewidth}
\centering
\includegraphics[scale=0.5]{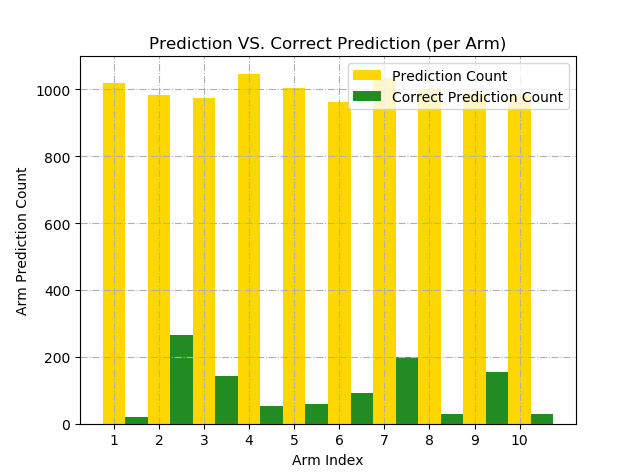}
\end{minipage}%
}%
\centering
\caption{ Mean UCB and Total VS. Correct prediction for
Alpha = $0.001/(\frac{correct-predictions}{10})$}
\end{figure}  
In the experiment, we further perform an $\alpha$ value with correct prediction dependent which is the Figure 4.8. The intuition for doing so is that the alpha value related to the correct prediction increases the exploitation of the arm that performs better (i.e., the larger the correct prediction, the smaller the $\alpha$ value). On the other hand, the number of explorations for the arm with inaccurate predictions has increased.

From the above figures, we can see that the arm with the largest upper confidence bound is also the arm with the highest number of predictions. This reflects the nature of the LinUCB algorithm that selects the arm with the highest upper confidence bound as the prediction.
\section{Matrix Factorization Implementation}
To implement the matrix factorization mentioned in section 3.1, a smaller data set was chosen in order for us to carry out a comparison between the two approaches: the base method and the bias method (i.e., the method of matrix factorization that considers both user bias and item bias). 

\textbf{\textit{Dataset}}. The partial dataset is shown in Figure 4.9. The user-item rating matrix $\textit{R}$ is a $25 \times 100$ matrix, which represent 25 users and 100 items. Furthermore, we can see from the figure that  $\textit{R}$ is a sparse matrix (i.e. matrix with many zeros)
\begin{figure}
\centering
\includegraphics[scale =0.8]{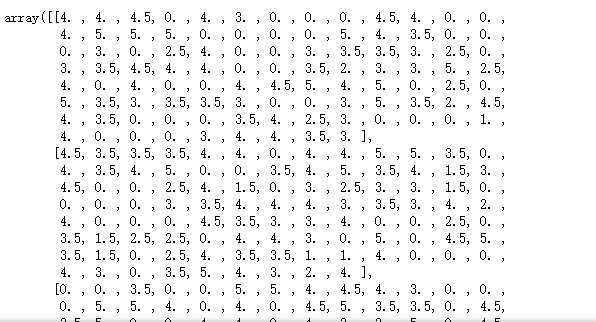}
\caption{Partial  of the user-item rating matrix \textit{R}}
\end{figure}

\textbf{\textit{Training}}.
Recall our loss function:
\begin{equation}
        \text { Loss }: J(p, q)=\min_{ q *, p * } \sum_{(u, i) \in K}\left(r_{u,i}-q_{i}^{T} p_{u}\right)^{2}+\lambda\left(\left\|q_{i}\right\|+\left\|p_{u}\right\|\right)^{2}
\end{equation}
Based on the methods proposed in \cite{mf1}, we use a stochastic gradient descent (SGD) algorithm to train two latent vector matrices: a user latent matrix \textit{p} and item latent matrix \textit{q}.By defining $e_{u,i}=r_{u,i}-q_{i}^{T} p_{u}$. We update the two matrices by the following formulas:
\begin{equation}
    q_{i} \leftarrow q_{i}+\eta\left(e_{u,i} p_{u}-\lambda q_{i}\right)
\end{equation}
\begin{equation}
    p_{u} \leftarrow p_{u}+\eta\left(e_{u,i} q_{i}-\lambda p_{u}\right)
\end{equation}
In the experiment, we set the number of latent features $k=2$, then we initiated the user matrix and item matrix with a size of $25 \times 2$ (i.e., number of users $\times$ number of latent features).
Furthermore, the learn rate$\eta=0.001$, regularization factor$\lambda = 0.1$ and iteration 1000 times.
\begin{figure}[htbp]
\centering
\subfigure[User latent matrix]{
\begin{minipage}[t]{0.5\linewidth}
\centering
\includegraphics[scale=0.6]{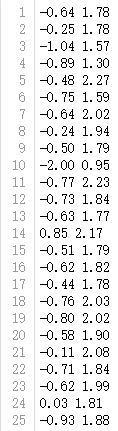}
\end{minipage}%
}%
\subfigure[Item latent matrix (first 25 out of 100 items)]{
\begin{minipage}[t]{0.5\linewidth}
\centering
\includegraphics[scale=0.6]{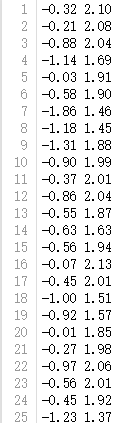}
\end{minipage}%
}%
\centering
\caption{User and item matrix}
\end{figure} 
After iteration, we can get our user latent matrix \textit{p} and item latent matrix \textit{q}, as shown in figure 4.10. Then, by performing the inner product between these two latent matrices, we get the predicted user-item rating matrix $\hat{R}$, shown in Figure 4.11. Expressed by the formula: $\hat{R}$ = user latent matrix $\times$ item latent matrix$^\textbf{T}$.
\begin{figure}[htbp]
\centering
\includegraphics[scale =0.7]{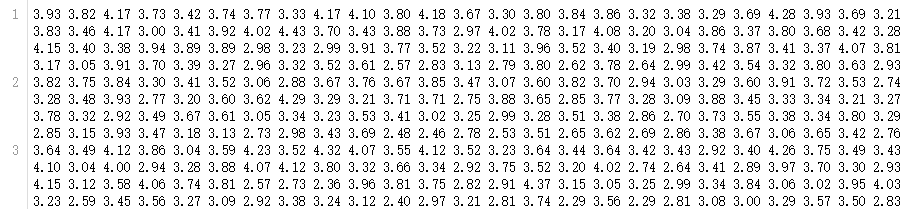}
\caption{Predicated rating matrix $\hat{R}$ (Partial,first 3 out of 100 user's estimate rating)}
\end{figure}

\textbf{\textit{Adding bias}}.Now, we will experiment with the bias method, which considers both user bias and item bias. As mentioned before, when we consider the bias, then the loss function becomes:
\begin{equation}
    \operatorname{Loss}=\min _{q^{*}, p^{*}} \sum_{(u, i) \in K}\left(r_{u,i}-b_{u,i}-q_{i}^{T}p_{u}\right)^{2}+\lambda \left(\left\|q_{i}\right\|^{2}+\left\|p_{u}\right\|^{2}+b_{u}^{2}+b_{i}^{2}\right)
\end{equation}
The formula used to iterate over the user latent matrix and the item latent matrix becomes:
\begin{equation}
    q_{i} \leftarrow q_{i}+\eta\left(\left ( r_{u i}-q_{i}^{T} p_{u}-b_{u,i} \right ) p_{u}-\lambda q_{i}\right)
\end{equation}
\begin{equation}
    p_{u} \leftarrow p_{u}+\eta\left(\left ( r_{u i}-q_{i}^{T} p_{u}-b_{u,i} \right ) q_{i}-\lambda p_{u}\right)
\end{equation}
The rest of the steps are consistent with the base matrix factorization method.After the training, we get the user and item latent matrix based on the bias method. By the visualization method, Figure 4.12 plots the latent vectors of users and items in the two-dimensional coordinate. Through multiply the red dot coordinate by the blue cross’s coordinate in the figure, you can get the predicted rating score of user $u$ on item $i$.

\begin{figure}[htbp]
\centering
\includegraphics[scale =0.8]{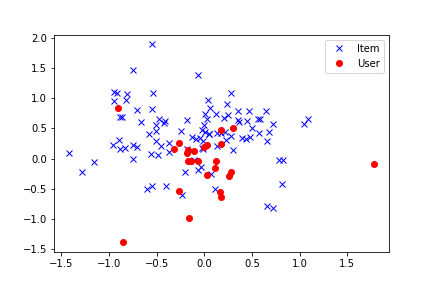}
\caption{visualization of user and item latent vector}
\end{figure}

\textbf{\textit{Evaluation}}. To compare the performance between the base method and the bias method, Mean Square Error (MSE) is used to evaluate these two algorithms. In order not to lose generality, we conducted 10 experiments and took the average of the results. What is striking in Table 4.1 is the dramatic decline of MSE, when we consider user bias and item bias, the value of MSE is reduced by $20\%$ which proved that consider both user and item bias when
perform the matrix factorization is effective.

\begin{table}[htbp]
\centering
\begin{tabular}{lll}
\hline
            &\textbf{MSE}    & \textbf{Improvement rate} \\ \hline
Base method  $\hat{r}_{u,i}=q_{i}^{T}p_{u}$ & 0.6530 &                  \\
Bias method $\hat{\boldsymbol{r}}_{u,i}=b_{u,i}+q_{i}^{T} p_{u}$ & 0.5178 & 20.715\%        
\end{tabular}
\caption{Evaluation between base method and bias method}
\end{table}
\section{Collaborative Filtering Recommender System: User-Based }
In this chapter, we will demonstrate the implementation of a user-based recommender system.

\textbf{\textit{Dataset}}. In this system, we adopt the dataset called MovieLen \cite{movielens}. MovieLen contains 100,000 ratings and 3,600 tag applications applied to 9,000 movies by 600 users.

\textbf{\textit{Implementation}}. By \textit{pandas} library, we read movies and ratings into the data frame. The format of both movie and rating data frames is shown in Figure 4.13.
    \begin{figure}[htbp]
\centering
\subfigure[movies data-frame]{
\begin{minipage}[t]{0.5\linewidth}
\centering
\includegraphics[scale=0.6]{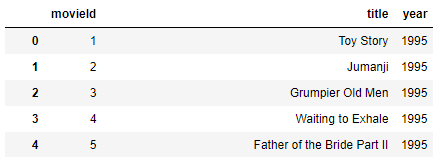}
\end{minipage}%
}%
\subfigure[ratings data-frame]{
\begin{minipage}[t]{0.5\linewidth}
\centering
\includegraphics[scale=0.7]{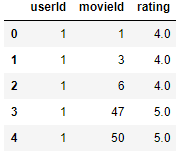}
\end{minipage}%
}%
\centering
\caption{Format of data frame}
\end{figure} 

According to the process for creating memory-based collaborative filtering, the next step is to identify a target user who will be recommended.

\textbf{\textit{Identify a target user}}. We define a target user and their rating in the way shown in the frame below. As in the image shown in Figure 4.14, the target user data frame involved the movie title with its ID and the target user’s rating. 
\begin{framed} 
\textbf{Target user input}\\
\rule{\textwidth}{0.1mm}
targetUserInput = [\\
{'title':'Grumpier Old Men', 'rating':4},\\
{'title':'Get Shorty', 'rating':2.5},\\
{'title':'Pulp Fiction', 'rating':3},\\
{'title':"Heat", 'rating':3.5},\\
{'title':'Toy Story', 'rating':5}\\
         ] \\
targetUserMovies = pd.DataFrame(targetUserInput)\\
targetUserMovies\\
\end{framed} 

\begin{figure}[htbp]
\centering
\includegraphics[scale =1]{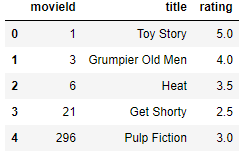}
\caption{Target user data frame}
\end{figure}

\textbf{\textit{Find similar user}}. With the movie ID in the target user input, we can now figure out the users who have seen and rated the movie, which is the same as the target user input. Then, according to the user ID, we group up the rows of the similar user sets. We sort these groups, with the rule that the group with more number of the movies watched or rated the same as the target user will be given a higher priority. With the information contained in Figure 4.15, the groups with user IDs 32, 91, and 414 are the groups that have the top 3 highest numbers of movies watched or rated the same as the target user.

\begin{figure}[htbp]
\centering
\includegraphics[scale =0.7]{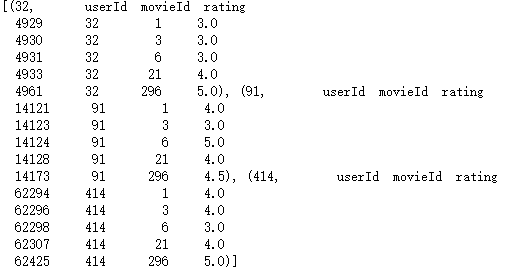}
\caption{Top three similar groups}
\end{figure}

\textbf{\textit{Calculate the similarity}}. In this step, we are going to compare all groups with the target user to figure out the most similar one by calculating the similarity by formula. As mentioned in the methodology chapter, there exist many formulae to calculate the similarity, but in this system, \textit{Pearson Correlation Coefficient} is adopted, and it can be represented as :
\begin{equation}
    \operatorname{sim}(i, j)=\frac{\sum_{p \in P}\left(R_{i, p}-\bar{R}_{i}\right)\left(R_{j, p}-\bar{R}_{j}\right)}{\sqrt{\sum_{p \in P}\left(R_{i, p}-\bar{R}_{i}\right)^{2}} \sqrt{\sum_{p \in P}\left(R_{j, p}-\bar{R}_{j}\right)^{2}}}
\end{equation}
where $R_{i, p}$ denote the rating of the user $i$ on movie $p$ and $\bar{R}_{i}$ represents the average rating of all movies by the user $i$. Similarly, $R_{j, p}$ denote the rating of the target user $j$ on movie $p$ and $\bar{R}_{j}$ represents the average rating of all movies by the target user $j$. Also, $P$ represents the set of all movies.

\begin{figure}
    \centering
    \includegraphics[scale=0.6]{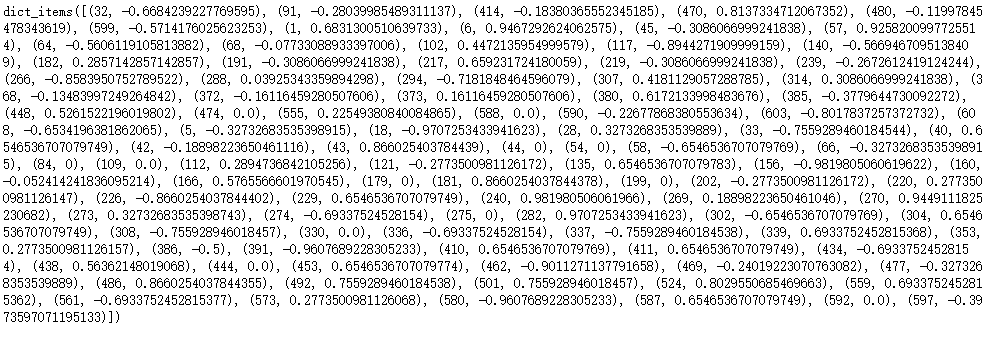}
    \caption{Similarity value of all similar users to the target user}
\end{figure}
After calculating the similarity, we can list out the similarity of all similar users to the target user, as shown in Figure 4.16.
\begin{figure}
    \centering
    \includegraphics[scale=1]{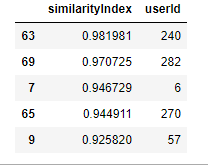}
    \caption{Top 5 similar users }
\end{figure}
As Figure 4.17 shows, by sorting the similar users by similarity, the user with ID 240 is the most similar user with the target user, for whom the similarity is 0.981981.

\begin{figure}[htbp]
\centering
\subfigure[Weighted rating]{
\begin{minipage}[t]{0.33\linewidth}
\centering
\includegraphics[scale=0.6]{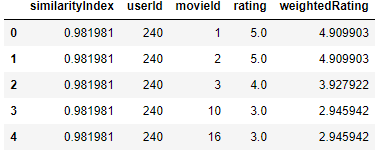}
\end{minipage}%
}%
\subfigure[Sum of the weighted rating]{
\begin{minipage}[t]{0.33\linewidth}
\centering
\includegraphics[scale=0.6]{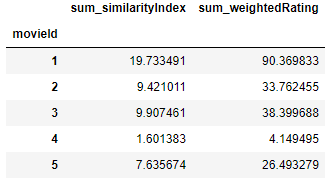}
\end{minipage}%
}%
\subfigure[Weighted average score]{
\begin{minipage}[t]{0.33\linewidth}
\centering
\includegraphics[scale=0.6]{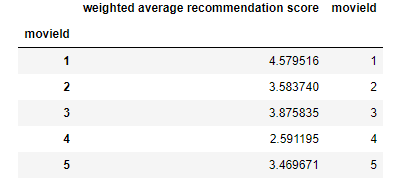}
\end{minipage}
}%
\centering
\caption{Calculate the score for recommendation (5 out of 9000 movies)}
\end{figure}
By the method of user-based collaborative filtering, we multiply the similar user’s rating by the similarity to get the weighted rating. As shown in Figure 4.18 (a), the intuition here is that the higher the similarity with the target user, the higher the weight of the target user's prediction score. Figure 4.18 (b) demonstrates the result of the sum of weighted ratings and the sum of similarity. Finally, in Figure 4.18 (c), by dividing the sum of weighted ratings by the sum of similarity, we can get the predicted weighted average recommendation score of what the target users probably rate the movie.

\textbf{\textit{Recommendation}}. For the recommendation, based on the above steps, first, sort the movies by average recommendation score, which is shown in Figure 4.19 (a). Then, by the movie ID in the sort list, we can get the top 10 movies with titles recommended by the user-based collaborative filtering system to the target user, as shown in Figure 4.19 (b).

 \begin{figure}[htbp]
\centering
\subfigure[Sorted by average recommendation score]{
\begin{minipage}[t]{0.5\linewidth}
\centering
\includegraphics[scale=0.7]{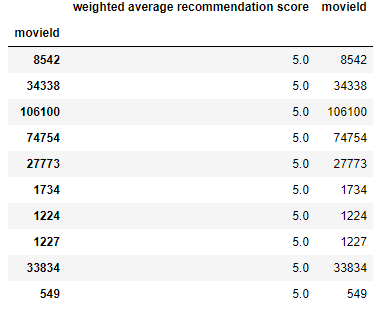}
\end{minipage}%
}%
\subfigure[recommend movies]{
\begin{minipage}[t]{0.5\linewidth}
\centering
\includegraphics[scale=0.7]{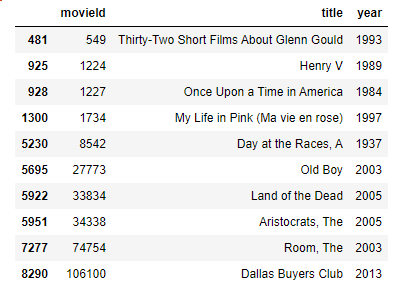}
\end{minipage}%
}%
\centering
\caption{Top 10 movies recommend by user-based system }
\end{figure}
\section{Hybrid Recommender System}
In this section, we will demonstrate the implementation of a hybrid system that combines matrix factorization and the content-based approach.

\textbf{\textit{Dataset}}. To verify the generality of the system, in addition to the MovieLens dataset we have used in the user-based system, we will also use a dataset called Jester\footnote[4]{Dataset available at: http://eigentaste.berkeley.edu/dataset} which contains over 1.7 million continuous ratings (-10.00 to +10.00) of 150 jokes from 59,132 users.

\textbf{\textit{Recommendation}}. Based on the method and implementation of matrix factorization described above, we can get our user latent matrix \textit{p} and item latent matrix \textit{q}. By performing the inner product between the above two latent matrices, we get the predicted user-item rating matrix $\hat{R}$. For the example shown in Figure 4.20, if we want to get the predicted rating of user 1 on item 1, we can use the inner product between two matrices to get the predicted rating. The process is shown in (a), and the predicted rating here is 3.999. When compared with the grand true rating shown in (b), we can see that the approximation of the predicted rating to the true rating is 99.975$\%$.

\begin{figure}[htbp]
\centering
\subfigure[Predicted rating]{
\begin{minipage}[t]{0.5\linewidth}
\centering
\includegraphics[scale=0.6]{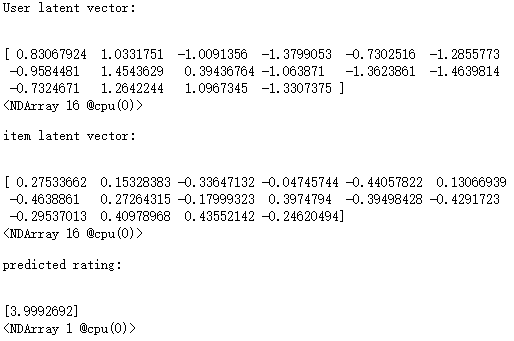}
\end{minipage}%
}%
\subfigure[Grand true rating]{
\begin{minipage}[t]{0.5\linewidth}
\centering
\includegraphics[scale=1]{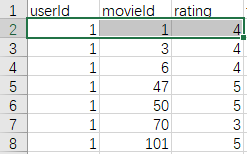}
\end{minipage}%
}%
\centering
\caption{comparison between predicted rating and true rating }
\end{figure} 

As per the methods that we talk about in section 3.3, the next step is to sort the movies that have not been rated by the target user through our predicted rating score and recommend the top-ranked movie to the target user. Through Figure 4.21, we can get the information that a movie with ID 910 is the movie with the highest predicted rating. 
\begin{figure}[htbp]
\centering
\includegraphics[scale =1]{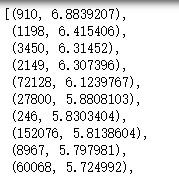}
\caption{Top 10 movies with high predicted rating}
\end{figure}

\textbf{\textit{Content-based recommendation}}. By the movie ID, we know that this movie is called \textit{``Some Like It Hot (1959)''}. All that is left for the hybrid recommendation system to do is to suggest movies similar to the movie \textit{``Some Like It Hot (1959)''} to the target user. 

Like what we described before, in order to measure how similar the movies are in attractiveness to the target user, we need to calculate the similarity by some formula. To test the ability of different similarity formulas, different from the \textit{Pearson Correlation Coefficient} used in the previous user-based system, we use \textit{Cosine similarity} to calculate the similarity, where the following formula can be used to calculate the similarity between movie $i$ and movie $j$:
\begin{equation}
\operatorname{similar}(i, j)=\cos (i, j)=\frac{i \cdot j}{\|i\| \cdot\|j\|}
\end{equation}
Figure 4.22, (a) shows the top 10 movies that are similar to the movie \textit{``Some Like It Hot (1959)''}. In (b), if we want to recommend two more movies to the target user, as the recommendation result of this hybrid system, movie \textit{`` The Fate of the Furious (2017)’’} which ID is 170875 and movie \textit{`` Withnail $\&$ I (1987)’’} which ID is 1202 will be recommended to the target user who likes the movie \textit{``Some Like It Hot (1959)''}.

\begin{figure}[htbp]
\centering
\subfigure[Top 10 items with high similarity]{
\begin{minipage}[t]{0.5\linewidth}
\centering
\includegraphics[scale=1]{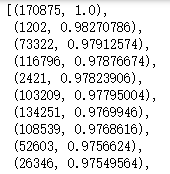}
\end{minipage}%
}%
\subfigure[Recommendation result]{
\begin{minipage}[t]{0.5\linewidth}
\centering
\includegraphics[scale=0.65]{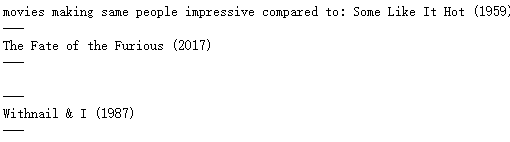}
\end{minipage}%
}%
\centering
\caption{Content-based recommendation}
\end{figure} 

\begin{figure}[htbp]
\centering
\includegraphics[scale =0.47]{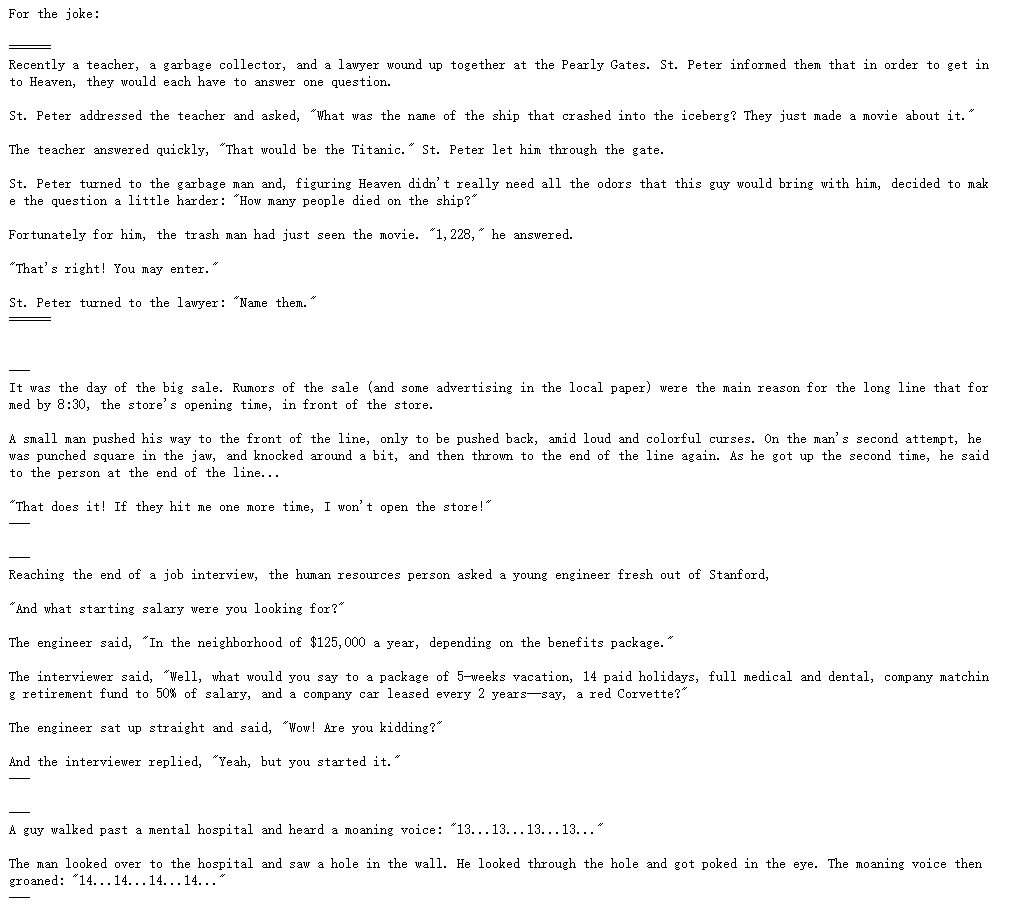}
\caption{Joke recommendation}
\end{figure}

When we repeated the above steps on another dataset, Jester on Jokes, our results are shown in Figure 4.23. By matrix factorization, we get the most recommended jokes for the target user (bounded by double lines), and based on content-based recommendations, three jokes similar to the matrix factorization recommended jokes are derived (bounded by signal line).
\section{BanditMF}
In this section, we will implement the BanditMF proposed in Section 3.4 and compare the experimental results obtained using different multi-armed bandit algorithms according to different criteria. In this experiment, the \textit{$\epsilon$-greedy}, \textit{UCB}, and \textit{Thompson sampling} algorithms will be used.We will implement BanditMF's process one by one as follows.

\textbf{\textit{Dataset}}. MovieLens will be used in this experiment, which contains the same data as we described in the last two experiments.

\textbf{\textit{Matrix factorization}}.  For BanditMF we use the bias method to implement matrix factorization, where $\hat{\boldsymbol{r}}_{u,i}=b_{u,i}+q_{i}^{T} p_{u}$ and $b_{u,i}$ is the combination of user bias, item bias, and average rating.

For the result shown in Figure 4.24, we generate a predicted user-item rating matrix with a size of 610 $\times$ 9724 
\begin{figure}[htbp]
    \centering
    \includegraphics[scale=0.8]{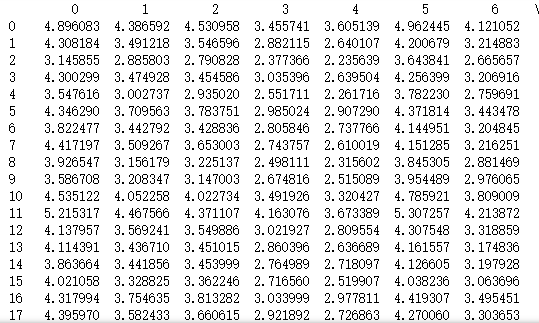}
    \caption{Partial of user-item predicted matrix }
\end{figure}

\textbf{\textit{Clustering}}. We perform the clustering on the row of the predicted user-item rating matrix $\hat{R}$, which means that we cluster the users who share similar rating preferences. In this experiment, we adopt a machine learning algorithm library called \textit{scikit-learn}. Based on the library, \textit{K-means} algorithm is adopted, which is measured by \textit{Euclidean distance}. We set the parameter \textit{n\_clusters}$=$3, which means that we will cluster all users into three clusters. Also, we set the value of parameter \textit{n\_init} to 20, which means that the \textit{K-means} algorithm will be run with 20 different centroid seeds and output the best result in 20 consecutive runs as the final result. Figure 4.25 shows the above iterations. As the number of iterations increases (i.e., x-axis), the sum of the Euclidean distances decreases (i.e., y-axis).

\begin{figure}[htbp]
    \centering
    \includegraphics[scale=0.6]{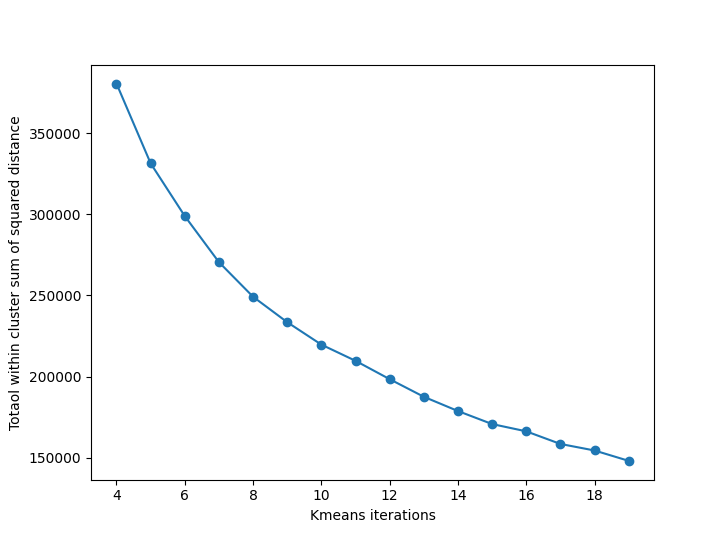}
    \caption{Iterations of K-means}
\end{figure}

Finally, by clustering, we got a 3 $\times$ 9724 predicted user-item rating matrix, shown in Figure 4.26, which means that we converted the original 610 $\times$ 9724 predicted user-item rating matrix to the 3 $\times$ 9724 clustered user-item rating matrix.
\begin{figure}[htbp]
    \centering
    \includegraphics[scale=0.6]{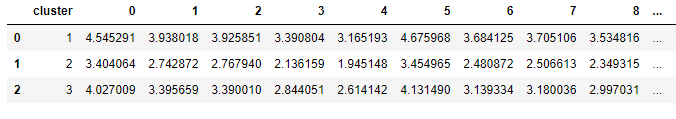}
    \caption{Predicted user-item rating matrix after clustering}
\end{figure}

\textbf{\textit{Online recommendation module}}.In the online recommendation, we used three different multi-armed bandit algorithms for the comparison, including \textit{$\epsilon$-greedy}, \textit{UCB}, and \textit{Thompson sampling}. The bandit algorithm takes the clustered predicted user-item rating matrix as data, and then selects and recommends items to the target users. After receiving the user's rating, the reward and regret are calculated. Then judge whether the user is still a cold user. If so, continue to update the Bandit algorithm with the received data; if not, transfer the user to the offline system and append the user-item rating matrix.

\textbf{\textit{Evaluation criteria}}. In addition to the \textit{rewards} and \textit{regrets} which are the common criteria in the multi-armed bandit algorithm, we have introduced two new criteria: \textit{discounted cumulative gain} (DCG) and \textit{normalized discounted cumulative gain} (NDCG) \cite{ndcg}. Their formulas are shown below:
\begin{equation}
    D C G(u)=r_{u, 1}+\sum_{t=2}^{T} \frac{r_{u, t}}{\log _{2} t}
\end{equation}
\begin{equation}
    N D C G=\frac{1}{N} \sum_{u} \frac{D C G(u)}{D C G^{*}(u)}
\end{equation}
We can see from the formula that these two criteria will discount the reward at a later stage, which means that the accuracy (i.e., user satisfaction) of early recommendations is more important than that of late recommendations, which is why these two criteria are applied here. Since we focus on solving the user's cold start problem, we need the algorithm to make accurate recommendations in a shorter period of time. However, when the user is no longer a cold user, accurate recommendations lose value because the user may give up using the server before you can make accurate recommendations.

\textbf{\textit{Result}}. We use $T$ to denote the total rounds, and $N$ denotes the number of new users. Further, $R_{t}$ represent the cumulative regret at $t$ rounds.
\begin{table}[htbp]
\centering
\begin{tabular}{llllll}
\hline
                  & $R_{1}$ & $R_{2}$ & $R_{3}$ & $R_{4}$   & $R_{5}$   \\ \hline
\textbf{Thompson sampling} & 0.00738798               & 0.01144652 & 0.04573369 & 0.08002087 & 0.11430805 \\
\textbf{UCB }              & 0.30009534 &0.61558008& 0.93106481& 1.23116015& 1.53125549 \\
\textbf{$\epsilon$-greedy }& 0.03761663 &0.07190381& 0.10619099& 0.14047816& 0.17476534
\end{tabular}
\caption{Cumulative regret when $T=5$, $N=2$}
\end{table}

\begin{table}[htbp]
\centering
\begin{tabular}{ll}
\hline
                  & \textbf{Normalized discounted cumulative gain (NDCG) } \\\hline
\textbf{Thompson sampling} & 0.974304261691225                                            \\
\textbf{UCB}          & 0.6949893798084221                                            \\
\textbf{$\epsilon$-greedy } & 0.9646494155015297        
\end{tabular}
\caption{NDCG for three algorithms when $T=5$, $N=2$}
\end{table}

From Table 4.2, we can see that Thompson sampling achieves the smallest regret, followed by $\epsilon$-greedy, while the UCB algorithm has the largest regret. Similarly,as shown in Table 4.3, Thompson sampling performs the highest NDCG which means it took less time to get the new user's preferences and provide a more accurate recommendation. The gap in NDCG between Thompson sampling and $\epsilon$-greedy is small, but for UCB the NDCG value is low, which means UCB cannot quickly get user preferences and make accurate recommendations.

\begin{table}[htbp]
\centering
\begin{tabular}{llllll}
\hline
                  & Cumulative regret & NDCG   \\ \hline
\textbf{Thompson sampling} &0.60372807               &  0.9588463913680102  \\
\textbf{UCB }              & 3.62539861 &0.7600041518097639 \\
\textbf{$\epsilon$-greedy }& 0.55885873 &0.9627427515312132
\end{tabular}
\caption{Cumulative regret and NDCG when $T=15$, $N=2$}
\end{table}
Table 4.4 shows the cumulative regret and NDCG when $T = 15$. After 15 rounds of recommendations, we see that the UCB algorithm also has the highest regret, but the difference is that $\epsilon$-greedy achieves a lower regret than Thompson sampling. Using the NDCG values, we conclude that $\epsilon$-greedy and Thompson sampling can make accurate recommendations faster compared to the UCB algorithm.
\begin{table}[htbp]
\centering
\begin{tabular}{llllll}
\hline
                  & Cumulative regret & NDCG   \\ \hline
\textbf{Thompson sampling} &1.36048677               &   0.9734318020634418  \\
\textbf{UCB }              & 15.49012907 &0.690944770845324 \\
\textbf{$\epsilon$-greedy }& 1.58943474 &0.9683140256825828
\end{tabular}
\caption{Cumulative regret and NDCG when $T=50$, $N=2$}
\end{table}

\begin{table}[htbp]
\centering
\begin{tabular}{llllll}
\hline
                  & Cumulative regret & NDCG   \\ \hline
\textbf{Thompson sampling} &0.27170972               &   0.9429132401340926  \\
\textbf{UCB }              & 1.36397039 & 0.7392537941258233 \\
\textbf{$\epsilon$-greedy }& 0.24749948 &0.9505001043510048
\end{tabular}
\caption{Cumulative regret and NDCG when $T=5$, $N=5$}
\end{table}
As Table 4.5 shows, when we increase $T$ to 50 rounds, Thompson sampling reverts to the algorithm that obtains the best results, both in terms of cumulative regret and NDCG, $\epsilon$-greedy achieves results comparable to Thompson sampling, and the gap between UCB and the two algorithms also exists.
In Table 4.6, when we increase the number of new users $N$ to 5, $T=5$, compared to the previous results in Table 4.2, although the UCB algorithm still performs the worst, the difference between it and Thompson sampling and $\epsilon$-greedy is slightly reduced. In Table 4.7, when we further set the number of new users $N$ to 50, the above conclusion still holds.In Table 4.8, $\epsilon$-greedy achieves a better performance than Thompson sampling when the number of users is relatively large. The regret of the UCB algorithm is 6.4 times higher than that of Thompson sampling.
\begin{table}[htbp]
\centering
\begin{tabular}{llllll}
\hline
                  & Cumulative regret & NDCG   \\ \hline
\textbf{Thompson sampling} &3.11502502               &   0.9378967672555214  \\
\textbf{UCB }              & 15.26485644 & 0.695220217000829 \\
\textbf{$\epsilon$-greedy }& 2.91051406 &0.9397737295609012
\end{tabular}
\caption{Cumulative regret and NDCG when $T=50$, $N=5$}
\end{table}

\begin{table}[htbp]
\centering
\begin{tabular}{llllll}
\hline
                  & Cumulative regret & NDCG   \\ \hline
\textbf{Thompson sampling} &4.77815387               &   0.9520246834804285  \\
\textbf{UCB }              & 30.58810434 & 0.6937173872467632 \\
\textbf{$\epsilon$-greedy }& 4.48261952 &0.9543088414431733
\end{tabular}
\caption{Cumulative regret and NDCG when $T=100$, $N=50$}
\end{table}

In summary, both Thompson sampling and $\epsilon$-greedy can achieve lower regret and higher NDCG. In most cases, Thompson sampling performs better, but the difference between the two is very small. In other words, users' preferences can be drawn and accurate recommendations can be made in a short time. As for the UCB algorithm, it performs the worst among the three algorithms and cannot quickly learn the preferences of new users. The possible reason is that the UCB algorithm needs to select all the arms once, a move that is fatal for the cold start problem. Because users will lose interest in the system if the system forces them to rate too much\cite{threshold}, it is important to get user preferences and make accurate recommendations in the shortest possible time for the cold start problem. However, if the number of new users is very large, e.g., 10,000 new users, then the UCB algorithm is equivalent to conducting 10,000 random selection strategies, which results in UCB not being able to make accurate recommendations for new users quickly and suffering from huge regret.
\chapter{Conclusion}
In this report, we first reviewed and formulated the representative multi-armed bandit and collaborative filtering methods. Subsequently, we introduce the matrix factorization methods, and a matrix factorization considering bias is proposed. Based on it, we illustrate the memory-based collaborative filtering recommender system and a hybrid recommender system that combines the content-based method with matrix factorization techniques. Importantly, in this work, BanditMF has been proposed, which is a multi-armed bandit-based matrix factorization recommender system. This system combines the matrix factorization (MF) which is model-based collaborative filtering with the multi-armed bandit algorithm. BanditMF uses a multi-armed bandit algorithm to solve the cold start problem in collaborative filtering. Also, with the help of matrix factorization results, the bandit algorithm can quickly derive user preferences. When we get the user preferences of the new users, we can transfer them to the offline module for the recommendation. That is, we add the user preference vector obtained from the online module to the user-item rating matrix. The advantage of this is that the collaborative filtering technology in the offline module fills a gap in the multi-armed bandit algorithm where similarity and user relationships are not taken into account. Also, for BanditMF, there are disadvantages. As we explain in the report, the contextual bandit algorithm can be based on contextual information to achieve a better balance between exploration and exploitation. However, in our BanditMF, we apply the context-free bandit, i.e., a bandit algorithm that does not consider contextual feature information, which may result in the recommended items not being optimal. The future work is to extend the contextual bandit to the BanditMF.

\printbibliography
\addcontentsline{toc}{chapter}{Bibliography}
\end{document}